\documentclass[fleqn,usenatbib]{mnras}

\usepackage{newtxtext,newtxmath}

\usepackage[T1]{fontenc}

\DeclareRobustCommand{\VAN}[3]{#2}
\let\VANthebibliography\thebibliography
\def\thebibliography{\DeclareRobustCommand{\VAN}[3]{##3}\VANthebibliography}

\usepackage{graphicx}	
\usepackage{amsmath}	
\usepackage{multirow}
\usepackage{physics}

\newcommand{\Baojiu}[1]{\textcolor{black}{#1}}
\newcommand{\Yuhao}[1]{\textcolor{black}{#1}}
\newcommand{\blue}[1]{\textcolor{blue}{#1}}
\newcommand{\Baojiunew}[1]{\textcolor{black}{#1}}
\newcommand{\Yuhaonew}[1]{\textcolor{black}{#1}}

\title[Nonlinear reconstruction of primordial features]{\Yuhao{Nonlinear reconstruction of features in the primordial power spectrum from large-scale structure} 
}

\author[Y.~Li, H.-M.~Zhu \& B.~Li]{
Yuhao Li,$^{1}$\thanks{E-mail: yl700@sussex.ac.uk (YL)}
Hong-Ming Zhu$^{2}$
and Baojiu Li$^{3}$
\\
$^{1}$Astronomy Centre, Department of Physics and Astronomy, University of Sussex, Falmer, Brighton, BN1 9RH, UK\\
$^{2}$Canadian Institute for Theoretical Astrophysics, University of Toronto, 60 St. George Street, Toronto, Ontario M5S 3H8, Canada\\
$^{3}$Institute for Computational Cosmology, Department of Physics, Durham University, South Road, Durham, DH1 3LE, UK
}

\date{Accepted 2022 June 1. Received 2022 June 1; in original form 2021 February 21}

\pubyear{2022}

\begin{document}
\label{firstpage}
\pagerange{\pageref{firstpage}--\pageref{lastpage}}
\maketitle

\begin{abstract}
{Potential} features in the primordial power spectrum
have been searched for in 
galaxy surveys in recent years since these features can assist in understanding the nature of inflation.
{The null detection to date suggests that any such features should be fairly weak, and next-generation galaxy surveys, with their unprecedented sizes and precisions, are in a position to place stronger constraints than before}. However, even if such primordial features once existed in the early Universe, they would 
{have been significantly} damped 
\Yuhao{in the nonlinear regime} at low redshift due to 
{structure formation}, which makes 
{them} difficult to be directly detected 
{in} real observations. 
{A potential way to tackle this challenge} for probing the features is to undo the cosmological evolution, i.e., {using} reconstruction {to obtain an approximate linear density field}. 
{By employing a \Yuhao{set} 
of} 
N-body simulations, here we show 
{that a recently-proposed} nonlinear reconstruction algorithm 
can effectively retrieve 
damped {oscillatory} features from 
\Yuhao{halo catalogues} 
and improve the accuracy of the 
{measurement} of feature parameters {(assuming that such primordial features do exist)}. 
{We do a Fisher analysis} to forecast 
{how} \Yuhao{nonlinear} reconstruction {affects the constraining power, and find that it can} lead to {significantly} more robust constraints on the 
{feature amplitude for a DESI-like} survey. 
\Baojiunew{Comparing \Yuhao{nonlinear} reconstruction with other ways of improving constraints, such as increasing the survey volume and range of scales, this shows that it is possible to achieve what the latter do, but at a lower cost.}

\end{abstract}



\section{Introduction}
Inflation, 
the most successful 
{theory to solve the problems of the hot Big Bang model and to explain the seeding of the observed large-scale structures today}, plays a crucial role in 
{the development} of modern cosmology. The 
{single-field} \Yuhao{slow-roll} inflation \citep{Guth:1980zm, Linde:1981mu, Albrecht:1982wi} predicts that primordial density fluctuations obey Gaussian statistics and the corresponding power spectrum follows a simple power law, which is 
favoured by the cosmic microwave background (CMB) data released by {the} WMAP 
\citep{Peiris:2003ff, Spergel:2006hy, Komatsu:2008hk, Hinshaw:2012aka} and Planck \citep{Planck:2013jfk, Ade:2015lrj, akrami2020} collaborations. 

However, the physical origin of the inflaton field\Baojiu{, which} is believed to have driven inflation\Baojiu{,} is not fully understood yet, and the fact that the very high energy \Baojiu{scale} in the early Universe makes it an ideal place to 
\Baojiu{probe the imprints} of the laws of fundamental physics offers 
\Baojiu{the possibility} that new physics can be revealed by cosmological observations of the large-scale structure \Baojiu{(LSS)}. 
\Baojiu{Certain} sophisticated models \Yuhao{of inflation and its alternatives} 
\Baojiu{developed over} the last decades 
\Yuhao{predict scale-dependent features in the power spectrum of primordial density fluctuations} 
\citep[see, e.g.,][for some reviews]{Bartolo:2004if, chen2010, Chluba:2015bqa, 2019BAAS...51c..98S}. 
\Baojiu{Such `feature models'} \Yuhao{can be mainly classified into three types 
\Baojiu{with} specific templates of oscillations added to the scale-invariant primordial power spectrum, each of which can be attributed to various mechanisms \citep[see, e.g.,][for some reviews]{chen2010, Chluba:2015bqa, 2019BAAS...51c..98S}. `Sharp-feature' models 
\Baojiu{have} sinusoidal wiggles \Baojiu{in the power spectrum, $P(k)$,} that oscillate linearly in wavenumber $k$ \Baojiu{at a fixed frequency, $\omega$}, which can be generated by a minimal local singularity such as a step in the inflationary potential 
\Baojiu{that} breaks the slow-roll condition \citep[e.g.,][]{1992JETPL..55..489S, Adams:2001vc, Chen:2006xjb, Hazra:2010ve, Adshead:2011jq, Hazra:2014goa}, or produced in particular cases of 
multi-field models of inflation \citep[e.g.,][]{Achucarro:2010da, Gao:2012uq}. Another 
type is \Baojiu{the} `resonant-feature' model whose oscillatory features are in logarithmic $k$, which can be realised in, e.g., the axion monodromy inflation \citep[][]{Flauger:2009ab, Flauger:2010ja}, or brane inflation \citep{Bean:2008na}, \Baojiu{models}. The last type is \Baojiu{the so-}called standard clock signal, which is a combination of the previous two feature models \citep[e.g.,][]{Chen:2011zf, Chen:2014joa, Chen:2014cwa}.}

\Baojiu{These feature models have been} continuously tested with the updated release of data from the Planck mission \citep{Ade:2013ydc, Ade:2015ava, Akrami:2019izv}, \Baojiu{but} none of them 
\Baojiu{has been found to be} preferable to the \Yuhao{scale-invariant power spectrum predicted by simple single-field slow-roll inflation models} so far, which suggests that such features should be fairly weak if they do exist. Since the primordial features are not only imprinted in the CMB, 
\Baojiu{but some of them} can also leave a signature in the matter \Baojiu{and galaxy distribution, future LSS} 
surveys, 
such as Euclid \citep{racca2016}, DESI \citep{aghamousa2016}, SPHEREx \citep{Dore:2014cca} and LSST \citep{LSST:2008ijt}, 
will provide the opportunity to search for, 
\Baojiu{or tighten the constraints on, them, complementary} to CMB data 
\citep[e.g.,][]{Huang:2012mr, Chen:2016vvw, Ballardini:2016hpi, Palma:2017wxu, LHuillier:2017lgm, Ballardini:2017qwq, Zeng:2018ufm}. \Yuhaonew{More recently, this idea has been put into practice by making forecast \citep[e.g.,][]{Beutler:2019ojk, Ballardini:2019tuc, Debono:2020emh} or performing real LSS data analysis \citep[][]{Beutler:2019ojk}.}

However, {any feature imprinted in the primordial density or curvature field by inflation is subject to the impact of cosmic evolution that 
\Baojiu{last until} today. In particular, even} if 
\Baojiu{such} primordial features once existed in the very early Universe, they would have been 
\Baojiu{modified} in the late-time Universe due to nonlinear structure formation. Meanwhile, the available information on large scales, where the evolution can be described by linear \Baojiu{perturbation} theory, is limited due to the cosmic variance\Baojiu{, i.e., the poor statistics caused by the finite number of Fourier modes probed in that regime}. This can affect the confidence level at which to measure or constrain these features. \Yuhao{In order to \Baojiu{maximally} extract useful information from the 
\Baojiu{observed galaxy distributions}, several studies of the primordial features in the nonlinear regime has been conducted. 
\Yuhaonew{\citet{Vasudevan:2019ewf,Beutler:2019ojk} analytically computed the damping effect by gravitational nonlinearities, making a considerable contribution to the forecast of constraints on primordial feature from future galaxy surveys. \citet{Ballardini:2019tuc} employed N-body simulations to show a compatible nonlinear damping effect with the analytic results above to leading order. \citet{Beutler:2019ojk} and \citet{Ballardini:2019tuc} made forecasts for future galaxy surveys by taking the damping effect into account. Besides, \citet{Beutler:2019ojk} performed the first LSS data analysis for the primordial features, which showed that LSS can surpass the CMB as a probe of such 
features.} Furthermore, \citet{Vlah:2015zda} and \citet{Chen:2020ckc} showed that different perturbation theories, including Lagrangian and Eulerian perturbation theories and  the effective field theory, can model the nonlinear evolution of primordial features well for $k \lesssim 0.25 \ h \rm Mpc^{-1}$ at $z=1$ and for $k \lesssim 0.2 \ h \rm Mpc^{-1}$ at $z=0$, but no oscillatory features survive past $k \approx 0.5 \ h \rm Mpc^{-1}$. Thus, it would be beneficial to \Baojiu{develop other approaches which can potentially allow us to} exploit the LSS data in the range of scales, $0.2\lesssim{k/(h{\rm Mpc}^{-1})}\lesssim0.5$ \Baojiunew{even at low redshifts.}} 

A potential method \Yuhao{mentioned in \citet{Vasudevan:2019ewf, Ballardini:2019tuc} and implemented in \citet{Beutler:2019ojk}} to address the issue \Yuhao{of nonlinear damping and further improve the constraints on primordial features} is to undo the cosmological evolution in a process 
called reconstruction, which can partially 
\Baojiunew{retrieve} the initial density field and \Baojiu{therefore} 
the information that 
existed \Baojiu{there.} 
A well-known example is the reconstruction of baryonic acoustic oscillation (BAO) features, which sharpens these features in the galaxy correlation function which provides a standard ruler for {distance measurements} \citep[e.g.,][]{Eisenstein:2006nk, kazin2014, Schmittfull:2015mja, Zhu:2016sjc, Wang:2017jeq, shi2018, sarpa2019, Mao:2020vdp}. \Yuhao{\Baojiu{While} reconstructing the primordial power spectrum from 
observed 
\Baojiu{galaxies} has been shown to be beneficial {for probing} the primordial features 
\Baojiu{from} LSS data \citep{Beutler:2019ojk}, \Baojiu{this study made use of one particular (the standard) reconstruction method, and it will be interesting to also assess how other reconstruction methods work in this regard.}} 

In this work, {as a first step towards assessing the potential benefit of \Yuhao{nonlinear} reconstruction,} we assume {additional simple oscillatory} features in the power-law primordial power spectrum. By utilising a \Yuhao{small set of} N-body simulations, we study the performance of 
\Baojiu{the} nonlinear reconstruction algorithm proposed {recently} by \citet{shi2018,Birkin:2018nag} \Baojiu{in retrieving} the damped primordial features from the 
\Yuhao{halo catalogues.} In particular, \Yuhao{by quantifying 
\Baojiu{this} damping 
\Baojiu{caused by structure formation} based on the functional form in \citet{Vasudevan:2019ewf, Beutler:2019ojk}}, we will carry out parameter fittings to the damped and reconstructed wiggles, \Baojiu{the comparison of which allows us} to assess whether nonlinear reconstruction can lead to more robust constraints on the feature parameters. To investigate the impact of \Baojiu{nonlinear} reconstruction in 
real galaxy surveys, we also forecast the constraints on the feature parameters for 
\Baojiu{a} DESI-like survey using the Fisher matrix approach, and compare the cases with and without reconstruction.

This paper is organised as follows: in Section~\ref{sec:2} we describe {the model of primordial features, the simulations used in this work, and the} methodology of assessing the performance of the nonlinear reconstruction method {to retrieve} the damped primordial features {due to structure formation}. In Section~\ref{sec:3} we give more details on the approach used to forecast the constraints on the feature parameters for the DESI-like survey. In Section~\ref{sec:4} we show the results of nonlinear reconstruction and forecast and discuss the implications of them. Finally, in Section~\ref{sec:5} we conclude our findings and {discuss potential} future improvements.

\section{Methodology}
\label{sec:2}
We start with presenting 
{the} primordial power spectrum models with oscillatory features {that we adopt in this paper for illustration purpose}. We then describe the simulation runs for these models. It is followed by a brief review of the nonlinear reconstruction method {which will be used to recover the small-scale oscillation features from evolved dark matter and halo fields}. Finally, we 
{describe the} analytic model to quantify the features measured in the power spectrum {before giving the details of the Fisher matrix forecast in the next section}.

\subsection{Models of featured primordial power spectrum}
\label{sec:2.1}
We take 
{a powerlaw-type} primordial power spectrum to be our fiducial no-wiggle model \Yuhao{(note that the BAO wiggles are still included)}, given by 
\begin{equation}\label{eq:2.1}
P_{\rm nw}^{\rm \Yuhao{ini}}(k) = A_{s} \bigg( \frac{k}{k_{\ast}} \bigg) ^ {n_{s} - 1},
\end{equation}
where $k$ is the comoving wavenumber, $A_s$ and $n_s$ are respectively the scalar amplitude and spectral index with the pivot scale given by $k_{\ast} = 0.05 \ \rm Mpc^{-1}$. 
\Yuhao{To explore whether the nonlinear reconstruction algorithm 
\Baojiu{employed in this paper can lead to improvements compared with} the unreconstructed cases in \citet{Ballardini:2019tuc}, we consider 
\Yuhaonew{four} wiggled models that are based on the template of the sharp feature model \citep{2019BAAS...51c..98S}, i.e., oscillations in linear $k$, given by}
\begin{equation}\label{eq:2.2}
P_{\rm w}^{\rm ini}(k) = P_{\rm nw}^{\rm ini}(k) \big[ 1 + A \Yuhaonew{\cos ( \omega k ^ m + \phi )} \big],
\end{equation}
where $A$, $\omega$ and $\phi$ are respectively the amplitude, frequency and phase of the oscillation. 
We extend the sharp feature model by introducing $m$ for a particular purpose explained later; when $m=1$, Eq.~(\ref{eq:2.2}) is related to the Eq.~(2.1) in \citet{Ballardini:2019tuc}.

Note that even if the primordial features exist, they could be more complicated than any phenomenological models that we are currently using. For now, we cannot determine the precise form of the features, thus we aim at something narrow, which is assuming that we know the functional form and verifying if \Yuhao{nonlinear} reconstruction can improve the accuracy of 
\Baojiunew{measuring} the feature parameters. 

The oscillation parameters of the 
\Yuhaonew{five} models are listed in Table~\ref{tab1}. \Yuhaonew{Note that the frequencies of the wiggled models here are in units of ${\rm Mpc}^m$ due to $m$ introduced above}. The initial oscillations of {the} 
\Yuhaonew{four} wiggled models are shown in the red dashed lines in the right panel of Fig.~\ref{fig:1}{, where we have presented the difference between $P_{\rm w}^{\rm \Yuhao{ini}}$ and $P_{\rm nw}^{\rm \Yuhao{ini}}$}. Within our 
{interested} range of scales, $k = (0.05-0.5) \ h \rm Mpc^{-1}$, \Yuhaonew{Model 1 has the first peak at the smallest scale, followed by Model 2 and Model 3, {the frequency used in Model 3 is the same as BAO frequency}. Model 4 is particularly adopted to have the first two peaks at the same positions of the first and third peaks of Model 2. The reason why this special model is designed will be explained in Section \ref{sec:4.2}.}
By comparing the reconstructed wiggles of the 
\Yuhaonew{four} wiggled models later, we would be able to 
comprehend the effect of the \Yuhao{nonlinear} reconstruction method on different scales.

\subsection{N-body simulations}
\label{sec:2.2}

{In the regime of linear perturbations, the primordial wiggles preserve their shapes and amplitude $P_{\rm w}^{\rm \Yuhao{ini}}/P_{\rm nw}^{\rm \Yuhao{ini}}$. However, nonlinear large-scale structure evolution will change this behaviour, leading to damping of $P_{\rm w}^{\rm \Yuhao{ini}}/P_{\rm nw}^{\rm \Yuhao{ini}}$ at late times. This makes it harder to measure the properties of these primordial oscillations \Baojiunew{directly} from an evolved density field, even more so for a late-time tracer (e.g., galaxy \Baojiunew{or halo}) field. In order to quantify such 
effects, N-body cosmological simulations can prove to be a useful tool.}

\begin{figure*}
	\includegraphics[width=1.95\columnwidth]{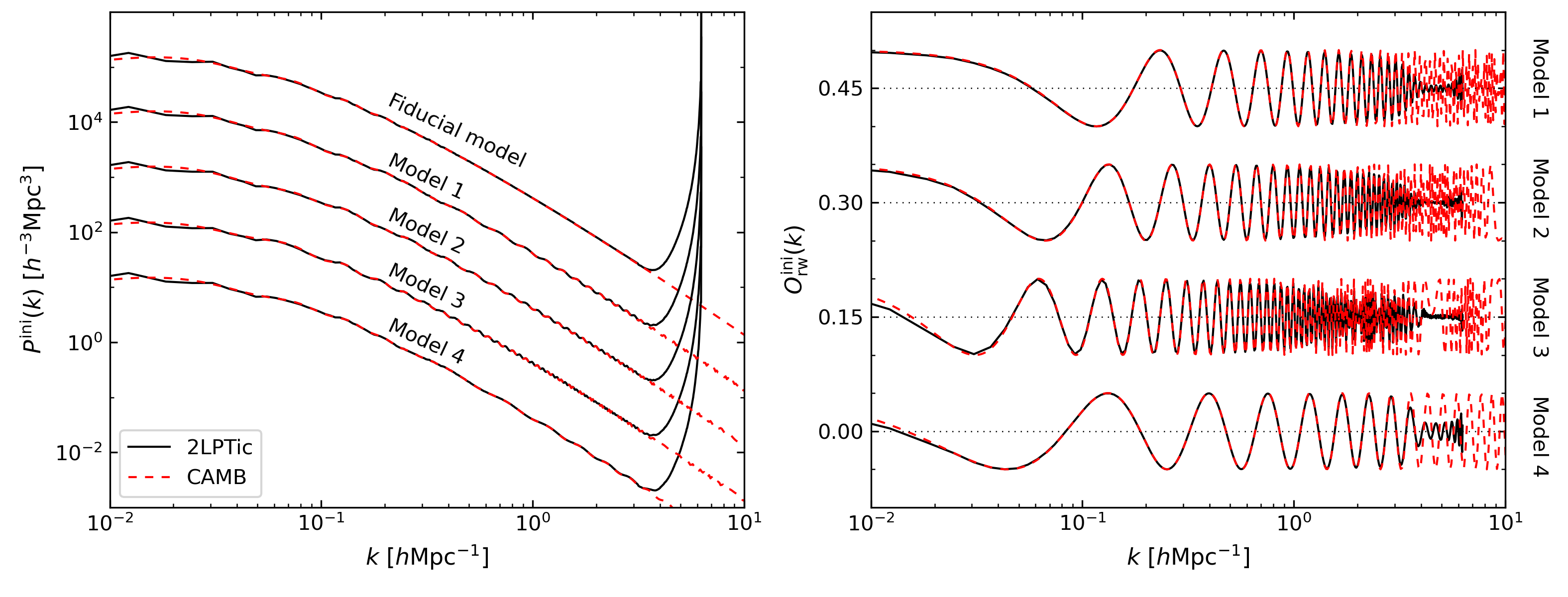}
    \caption{{[Colour Online]} The left panel shows the comparison between the initial matter power spectra given by 
    {{\sc camb}} (red dashed lines) and the matter power spectra measured from the initial conditions of the simulations generated using 
    {2{\sc lpt}ic} (black lines), from the top down they are respectively the fiducial model, Model 1, Model 2, Model 3 \Yuhaonew{and Model 4}, each model is shifted upwards by a factor of $10$ successively to avoid the clutter of all curves. The right panel shows the 
    {$\Yuhao{O}_{\rm rw}^{\rm ini}$ results, cf., Eq.~\eqref{eq:2.0},} obtained from the left panel for {the} four wiggled models, for instance, the top curve shows the ratio of Model 1 to the fiducial model, followed by the ones for Model 2, Model 3 \Yuhaonew{and Model 4} downwards; each model is shifted upwards by a constant of $0.15$ successively for the same reason {as above}.}
    \label{fig:1}
\end{figure*}

We have run \Yuhaonew{five} simulation runs including the no-wiggle model and \Yuhaonew{four} wiggled models. First we assume a flat universe and adopt Planck 2018 cosmology, with $h = 0.674$, $\rm \Omega_{m} = 0.3135$, $\Omega_c h^2 = 0.120$, $\Omega_b h^2 = 0.0224$, $\rm \Omega_{\Lambda} = 0.6865$, $n_{s} = 0.965$ and $A_s = 2\times 10^{-9}$ \citep{aghanim2020}. The value of $\sigma_8$ is approximately 0.79 
{though} it varies a little bit across different models. We then customise the function of the primordial power spectrum in the Einstein-Boltzmann solver code 
{\sc camb} \citep{lewis2011} to be Eq.~(\ref{eq:2.1}) for the no-wiggle model and 
Eq.~(\ref{eq:2.2}) for the wiggled models. We calculate the linear theory matter power spectrum at $z=49$ using 
{this version of the {\sc camb} code}, which is used as the input matter power spectrum for the publicly available code 
{2{\sc lpt}ic} \citep{crocce2006} to generate the initial conditions used for the N-body simulations. In the left panel of Fig.~\ref{fig:1} we compare the initial matter power spectrum given by 
{{\sc camb}} and the matter power spectrum measured from the initial conditions generated using {2{\sc lpt}ic; it can be seen that} they are in good agreement for all models within the range of scales of our interest {(the blowing up at small scales is due to the finite particle resolution).}

To {more} conveniently describe the oscillatory features for the wiggled models, {as mentioned above,} we define the relative wiggle 
{pattern} as  
\begin{equation}\label{eq:2.0}
\Yuhao{O}_{\rm rw}^{\rm \Yuhao{ini}}(k) = \frac{P_{\rm w}^{\rm \Yuhao{ini}}(k)}{P_{\rm nw}^{\rm \Yuhao{ini}}(k)}-1,
\end{equation}
which are shown in the right panel of Fig.~\ref{fig:1}. 
{This} 
indicates that the oscillatory features 
\Baojiu{have been reliably} created in the initial conditions of the simulations \Yuhao{within our interested range of scales, \Baojiunew{e.g., $k\lesssim0.5$ $h\mathrm{Mpc}^{-1}$}}. 

Next, we run the simulations using the {parallel} N-body code 
{{\sc ramses}} \citep{teyssier2002} which is based on the adaptive mesh refinement (AMR) technique. Each simulation is performed with $N = 1024^3$ dark matter particles in a box of size $1024 \ h^{-1} \rm Mpc$, {and} we output four snapshots at different redshifts, respectively as $z = 0$, $0.5$, $1$, and $1.5$. For each snapshot, we use the halo finder 
{{\sc rockstar}} \citep{behroozi2013} to identify the haloes with the definition of the halo mass $M_{200c}$, where $M_{200c}$ is the mass within a sphere whose average density is 200 times the critical density. Since the low-mass haloes are unable to be fully probed due to the limited simulation resolution, we measure the cumulative halo mass functions (cHMFs) from the 
{main} haloes with more than $100$ particles to check the validity of the simulation, which show very good agreement with the analytic formulae in \citet{tinker2008}. For each snapshot we establish one dark matter particle catalogue (hereafter DM) and two 
\Baojiunew{halo catalogues} respectively with the number density of $1 \times 10^{-3} (h^{-1} \rm Mpc)^{-3}$ (hereafter H1) and $5 \times 10^{-4} (h^{-1} \rm Mpc)^{-3}$ (hereafter H2). Both host haloes and subhaloes are included in the halo catalogues. \Yuhaonew{The number density of $5 \times 10^{-4} (h^{-1} \rm Mpc)^{-3}$ is chosen to be an approximate value according to the current observations such as CMASS or LOWZ despite not being exactly the same, and $1 \times 10^{-3} (h^{-1} \rm Mpc)^{-3}$ is a representative value of emission line galaxies (ELGs) in DESI survey; these choices are also somehow limited by the resolution of our simulations, though the use of the dark matter density field serves as a catalogue that has a much larger number density. Many realistic mock galaxy catalogues would give something between H1 and DM.}

We achieve the number density by applying a mass cutoff, i.e., 
{neglecting} the haloes with smaller masses than the cutoff. By using the power spectrum estimator tool 
{{\sc powmes}} \citep{colombi2011}, we measure the nonlinear matter power spectrum from DM and nonlinear halo power spectrum separately from H1 and H2. Finally, we take the ratio of the power spectrum of the wiggled models to 
{the} corresponding power spectrum of the no-wiggle model to obtain the 
{quantity $\Yuhao{O}_{\rm rw}$} for all cases.

\begin{table}
\centering
\caption{The oscillation parameters used for the no-wiggle model and \Yuhaonew{four} wiggled models. Columns respectively denote (1) the power of the comoving wavenumber; (2) the amplitude, (3) frequency and (4) phase of the oscillation.}
\label{tab1}
\begin{tabular}{c|c|c|c|c} 
\hline
& $m$ & $A$ & $\omega $ & $\phi/\pi$\\
&  &  & \Yuhaonew{$[{\rm Mpc}^{m}]$} & \\
\hline
Fiducial &  & $0$ &  &  \\
Model $1$ & $1$ & $0.05$ & \Yuhaonew{$40$} & $0$ \\
Model $2$ & $1$ & $0.05$ & \Yuhaonew{$70$} & $0$ \\
Model $3$ & $1$ & $0.05$ & \Yuhaonew{$150$} & $0$ \\
Model $4$ & $0.631$ & $0.05$ & \Yuhaonew{$28.9$} & $0$ \\
\hline
\end{tabular}
\end{table}
\subsection{Reconstruction}
\label{sec:2.3}
In order to {partially} 
\Baojiu{retrieve} the primordial features damped during 
{structure formation}, we 
{perform reconstruction of} the initial density field from the late-time density field using 
{the} 
nonlinear reconstruction algorithm 
{described in} \citet{shi2018}. This reconstruction method is based on mass conservation. 
Without assuming a cosmological model or {having} free parameters except the size of the mesh used to calculate the density field, it employs multigrid Gauss-Seidel relaxation to solve the nonlinear partial differential equation which governs the mapping between the initial Lagrangian and final Eulerian coordinates of particles in evolved density fields. 
Previous tests show that the reconstructed density field is over $\sim 80\%$ correlated with the initial density field for $k \lesssim 0.6 \ h \rm Mpc^{-1}$, {if reconstruction is performed on the dark matter density field,} which 
cover the scales of our interest, {but the performance becomes poorer when the method is instead applied on the density fields calculated from sparse tracers \citep{Birkin:2018nag,Wang:2019zuq,Liu:2020pvy}. This method is implemented in a modified version of the {\sc ecosmog} code \citep{li2012, li2013}, which itself is based on {\sc ramses}.}

We reconstruct the initial density field separately from the catalogues DM, H1 and H2 for each snapshot. 
The halo catalogues, {which contain both main and subhaloes,} are assumed to be the same as mock galaxy catalogues hereafter unless otherwise stated\footnote{\Baojiunew{As a result, we will use `haloes' and `galaxies' interchangeable throughout the rest of this paper: `galaxies' will be used where we refer to observational quantities, while `haloes' will be used for simulated quantities.}}. The procedure for the reconstruction from the halo catalogue is principally similar to that from the dark matter particle catalogue, 
{apart from} two things at the beginning. One is that we prepare the Gadget-format particle data for the 
{{\sc ecosmog}} code in two ways. The halo catalogue is directly written into Gadget-format tracer particles due to its small number density. However, the very large number 
of the simulation particles{, along with their strongly non-uniform spatial distribution,} in the dark matter particle catalogues, leads to the requirement of 
{large memory footprint} when processing the data. To avoid this problem, we use the publicly available 
{{\sc dtfe}} code \citep{cautun2011}, based on Delaunay tessellation, to calculate the density field on a regular mesh with $512^{3}$ cells employing the triangular shaped cloud (TSC) mass assignment scheme; then the mesh cells are regarded as 
{uniformly-distributed} fake particles with different masses, which are 
{transformed} to 
Gadget format {that can be directly read by {\sc ecosmog}.} 

The other particular thing is that we calculate the linear halo bias used for the reconstruction from the halo catalogue. The estimate of the halo bias is based on the relation
\begin{equation}\label{eq:2.3}
b_{1}(r) = \frac{\xi_{\rm hh}(r)}{\xi_{\rm hm}(r)},
\end{equation}
where $\xi_{\rm hh}(r)$ is the auto-correlation function of haloes and $\xi_{\rm hm}(r)$ is the cross-correlation function between the haloes and the dark matter particles. 
We use the publicly available 
{{\sc cute}} code \citep{alonso2012} to 
{measure} $\xi_{\rm hh}(r)$ and $\xi_{\rm hm}(r)$ from a given simulation snapshot, and take the ratio between them to obtain the value of linear halo bias as a function of the distance $r$. 
Since the linear halo bias is theoretically a constant on large scales, we apply the method of least squares to the values on scales $r \gtrsim 10 \ h^{-1} \rm Mpc$ to 
{obtain an} estimate {of it. Note that when dealing with observational data we do not necessarily have such an accurate measurement of the linear halo or galaxy bias; however, \citet{Birkin:2018nag} find that the exact value of linear bias is not very important for this reconstruction method to recover the phases of the initial density field.} 

The following steps of reconstruction are then the same for both dark matter particle catalogue and halo catalogues. First, 
{{\sc ecosmog}} 
calculates the density field in the Eulerian coordinates using the TSC mass assignment scheme, {and solves the mapping between the Eulerian and Lagrangian coordinates, to get} 
the displacement potential 
as well as the displacement field on a regular mesh with $512^{3}$ cells. We then use a 
Python code to transfer the output fields from the Eulerian coordinates to the Lagrangian coordinates. After that, {because the Lagrangian coordinates are not uniform,} we feed the 
{{\sc dtfe}} code with the Lagrangian coordinates and displacement field of the mesh cells to calculate the reconstructed density field as the divergence of the displacement field w.r.t. the Lagrangian coordinates. Finally, we measure the reconstructed power spectrum from the reconstructed density field using a post-processing code.

\subsection{
{Parameter fitting to the damped wiggles}}
\label{sec:2.4}

{As we discussed above, cosmic structure formation leads to damping of the primordial wiggles. Reconstruction is expected to revert some of this damping, but cannot completely undo it. So we need a model for the wiggles of the reconstructed matter or halo power spectrum. Ideally this should be an analytical model since it can be more easily used in the Fisher analysis later. In this subsection, we describe how this is achieved by using a fitting function.}

\Yuhao{A functional form of the feature damping is analytically computed in \citet{Vasudevan:2019ewf} and \citet{Beutler:2019ojk} to be a Gaussian. \Baojiu{We} combine it with 
\Baojiu{the} oscillatory feature model \Baojiu{described above, in order} to directly fit the wiggle pattern $\Yuhao{O}_{\rm rw}$. 
\Baojiu{The fitting function that is used to described the damped wiggles is given by}
}
\begin{equation}\label{eq:2.4}
{O}_{\rm rw}^{\rm fit}(k,z) =  A \Yuhaonew{\cos ( \omega k ^ m + \phi )} \exp \bigg[-\frac{k^{2}\zeta(z)^{2}}{2} \bigg],
\end{equation}
where $\zeta(z)$ is the damping parameter that depends on the redshift $z$. For the fitting of each 
{measured \Yuhao{$O_{\rm rw}(k)=P_{\rm w}(k)/P_{\rm nw}(k)-1$}}, we let $\omega$, $\phi$ and $\zeta$ be the free parameters because $\omega$ and $\phi$ play an essential role in determining the position of the peaks, and $\zeta$ quantifies the extent of the damping effect. The parameters $A$ and $m$ are taken to be their theoretical values in Table~\ref{tab1}. {In principle, $A$ is also a free parameter here and should be allowed to vary in our parameter fitting. We have explicitly checked this 4-parameter fitting and found that, compared with the 3-parameter fitting, in the vast majority of cases of Table \ref{tab2}, the best-fit values of $\omega$ and $\phi$ are not more accurate, which is as expected. There is a degeneracy between the amplitude $A$ and the damping scale $\zeta$, with the fitted values of the latter having larger uncertainties in the case of the 4-parameter fitting. Since for our forecast work the value of $\zeta$ is more important, we stick with the results obtained from the 3-parameter fitting.}

We apply the least-squares estimator to obtain the best-fit parameters by minimising
\begin{equation}\label{eq:2.5}
{\chi^2 =} \sum_{i=1}^{N} \big[\Yuhao{O}_{\rm rw,i}(z)-\Yuhao{O}_{\rm rw}^{\rm fit}(k_{\rm i}, z; \omega, \phi, \zeta) \big]^{2},
\end{equation}
where $\Yuhao{O}_{\rm rw,i}(z)$ are the data points of wiggle spectrum {in the $i$th $k$ bin at reshift $z$}. Since there is only one realisation of simulation for each model, we assume that the uncertainties of all data points {$\Yuhao{O}_{\rm rw,i}(z)$} are the same and follow the same Gaussian distribution. 
{Note that, as the quantity we fit is $\Yuhao{O}_{\rm rw}=P_{\rm w}/P_{\rm nw}-1$, this is equivalent to doing the fitting of $P_{\rm w}$ with $\sqrt{P_{\rm nw}}$ as uncertainty \citep[e.g.,][]{Feldman:1993ky}.}

We calculate the uncertainties of the best-fit parameters based on 95 \% confidence interval, 
{as} a rough estimate of the size of the errors. To 
{minimise} the influence of the 
{cosmic variance} on very large scales, we fit the data within the interval of $k = (0.04-0.6) \ h \rm Mpc^{-1}$, which 
covers our intended range of scales. 

\section{Forecast for the DESI-like survey}
\label{sec:3}
In order to investigate the impact of reconstruction, 
we will forecast the constraints on the feature parameters for the DESI-like survey using the Fisher information matrix{, and compare with the case of doing no reconstruction. For this purpose, we} first model the observed broadband galaxy power spectrum. Then we describe how to calculate the Fisher information matrix, followed by its analytic marginalisation. Finally, we 
{give} the specifications of the DESI-like survey.

\subsection{Modelling the observed galaxy power spectrum}
\label{sec:3.1}
Combining the Eqs.~(\ref{eq:2.0}) and (\ref{eq:2.4}), the featured 
\Baojiunew{nonlinear matter} power spectrum in real space can be modelled as, 
\begin{equation}\label{eq:4.1}
P_{\rm mod}(k,z) = P_{\rm nl}(k,z) \bigg[1 + A \Yuhaonew{\cos ( \omega k ^ m + \phi )} \exp \bigg(-\frac{k^{2}\zeta(z)^{2}}{2} \bigg) \bigg],
\end{equation}
where $P_{\rm nl}(k,z)$ is the nonlinear matter power spectrum without the {primordial} oscillatory features at $z$, which includes the BAO wiggles and is equivalent to the nonlinear matter power spectrum of the no-wiggle model. However, since there is only one {simulation} realisation for 
{a single} no-wiggle model, which cannot provide a smooth nonlinear matter power spectrum, {and since a fast method to get $P_{\rm mod}$ is more convenient in the Fisher analysis,} we use the 
{{\sc halofit}} model 
{in the {\sc camb}} code to calculate $P_{\rm nl}(k,z)$ instead \Baojiunew{later in this work}. \Yuhaonew{We have checked that the fractional difference between the simulated no-wiggle power spectrum and the one computed by {\sc halofit} is below $10\%$ within the entire fitting range.}

The broadband galaxy power spectrum in real space is not a direct observable due to the measurement in the angular and redshift coordinates instead of the 3D comoving coordinates. In order to relate the observed galaxy power spectrum $P_{\rm obs}(\boldsymbol{k},z)$ to the modelled 
\Baojiunew{matter} power spectrum $P_{\rm mod}(k,z)$, the standard practice is to project the galaxies to their comoving positions assuming some reference cosmology via the coordinate transformation based on the relations
\begin{equation}\label{eq:4.2}
k_{\perp}^{\rm ref} = \frac{D_{\rm A}(z)}{D_{\rm A}^{\rm ref}(z)}k_{\perp},
\quad
k_{\parallel}^{\rm ref} = \frac{H^{\rm ref}(z)}{H(z)}k_{\parallel},
\end{equation}
where {$k_{\parallel}$ and $k_{\perp}$} are respectively the 
\Yuhao{line-of-sight} and transverse components of the 
{wavevector $\boldsymbol{k}$}, i.e., $k^{2} = |\boldsymbol{k}|^2 = k_{\perp}^{2} + k_{\parallel}^{2}$, the superscript $^{\rm ref}$ denotes the reference cosmology, note that the reference cosmology hereafter is the same one used in the simulations unless otherwise stated; $D_{\rm A}(z) = r(z)/(1+z)$ is the angular diameter distance at $z$ with the comoving distance $r(z)$: under the assumption of flat universe it is given by
\begin{equation}\label{eq:4.3}
r(z) = \frac{c}{H_0}\int_{0}^{z} \dd z^\prime \Big[\Omega_{\rm m}(1+z)^3 +\Omega_{\Lambda}\Big]^{-\frac{1}{2}},
\end{equation}
{where $\Omega_\Lambda=1-\Omega_{\rm m}$ is the current density parameter of the cosmological constant,} and the Hubble parameter $H(z)$ is given by
\begin{equation}\label{eq:4.4}
H(z) = H_0 \Big[\Omega_{\rm m}(1+z)^3 +\Omega_{\Lambda}\Big]^{\frac{1}{2}}.
\end{equation}
Along with several main factors being considered, i.e., the redshift-space distortions (RSD) and shot noise, one can model the observed galaxy power spectrum as 
\begin{equation}\label{eq:4.5}
P_{\rm obs}(k,\mu,z) = \Bigg[ \frac{D_{\rm A}^{\rm ref}(z)}{D_{\rm A}(z)} \Bigg]^{2} \frac{H(z)}{H^{\rm ref}(z)} \frac{F_{\rm FoG}(k, \mu, z)}{\sigma_{8}^{2}(z)} P_{\rm mod}(k,z) + N_{\rm gal}(z),
\end{equation}
where $\sigma_{8}(z)$ is the {R.M.S.} linear density fluctuations on the scale of $8h^{-1}\rm Mpc$, $N_{\rm gal}(z) = 1 / \overline{n}_{\rm g}(z)$ is the shot noise with $\overline{n}_{\rm g}(z)$ being the galaxy number density, and the Finger-of-God factor $F_{\rm FoG}(k, \mu, z)$ describing the effect of RSD 
{is} modelled as \citet{Ballardini:2019tuc}
\begin{equation}\label{eq:4.6}
F_{\rm FoG}(k, \mu, z) = \frac{\big[b(z)\sigma_8(z) + f(z)\sigma_8(z)\mu^2\big]^2}{1 + k^2 \mu^2 \sigma_{r,p}^2 / 2} \exp \big(- k^2 \mu^2 \sigma_{r,z}^2 \big),
\end{equation}
where \Baojiunew{we have included the linear halo bias at $z$, $b(z)$, to make $P_{\rm obs}$ the `galaxy' (remember that in our simulations we treat (sub)haloes as mock galaxies) power spectrum, and}
\begin{equation}
    f(z) = \frac{\dd\ln{D(a)}}{\dd\ln{a}},
\end{equation} 
is the linear growth rate at $z$ with $D(a)$ and $a$ respectively being the linear growth factor and the scale factor (note that we normalise $D(a)$ so that $D(a=1) = 1$ 
{in this work}), $\mu=\cos{\theta}$ with $\theta$ being the angle between the 
{wavevector $\boldsymbol{k}$} and the line of sight, 
i.e., $\mu = k_{\parallel}/k$, $\sigma_{r,p} = \sigma_{p}/[H(z)a]$ is the distance dispersion corresponding to the physical velocity dispersion $\sigma_{p}$ whose fiducial value is taken to be $290 \rm \ km~s^{-1}$. \Yuhao{The {last} exponential factor 
represents an additional damping to account for the observational redshift error $\sigma(z)$ with $\sigma_{r,z} = c \sigma(z)/H(z)$ specific to a given survey, which is very close to 1 for our intended range of scales given that the DESI survey assumes $\sigma(z) = 0.0005 / (1 + z)$ \citep{aghamousa2016}, so we neglect it in the calculation.} 

Additional effects involved in 
\Baojiunew{real observational constraints,} such as the survey window function and finite bandwidths, \Yuhaonew{which would influence the forecasted constraining power to some extent \citep[see, e.g.,][for a more detailed discussion]{Beutler:2019ojk}}, should be taken into account when dealing with 
real surveys in future works, \Baojiunew{but these are not included in the} forecast \Baojiunew{here. The present work is therefore a simplified proof-of-concept study which is likely to lead to optimistic forecasts.}

\subsection{Fisher information matrix}
\label{sec:3.2}
The Fisher matrix approach provides a method to propagate the uncertainties of the observable to the constraints on the cosmological parameters. Our calculation of the Fisher matrix is based on \citet{tegmark1997} and \citet{seo2003}, assuming that the power spectrum of a given $k$ mode satisfies {a} Gaussian distribution {which has a variance equal to the power spectrum itself}, and that different 
{bins} of $k$ are independent of each other for large surveys, the Fisher matrix for each redshift bin, with bin centre at $z = z_{\rm c}$, can be approximated as
\begin{eqnarray}\label{eq:4.7}
&& F_{ij}(z_{\rm c})=\frac{V_{\rm eff}(z_{\rm c})}{4\pi^{2}} \int_{0}^{1} \dd \mu\nonumber\\ 
&& \times\int_{k_{\rm min}}^{k_{\rm max}} \dd k k^{2} \frac{\partial \ln{P_{\rm obs}(k,\mu,z_{\rm c})}}{\partial \theta_{i}} \frac{\partial \ln{P_{\rm obs}(k,\mu,z_{\rm c})}}{\partial \theta_{j}},
\end{eqnarray}
where {$k_{\rm min}, k_{\rm max}$ are respectively the minimum and maximum values of $k$ used for the forecast}. We set $k_{\rm min}=0.05 \ h\rm Mpc^{-1}$ and {adopt} two values of $k_{\rm max}$, respectively 
$0.25 \ h\rm Mpc^{-1}$ and $0.5 \ h\rm Mpc^{-1}$, to compare the constraints for different ranges of scales. The effective volume of the redshift bin $V_{\rm eff}(z_{\rm c})$ is expressed as
\begin{equation}\label{eq:4.8}
V_{\rm eff}(z_{\rm c}) = \bigg[1 + \frac{1}{\overline{n}_{\rm g}(z)P_{\rm obs}(k,\mu,z)} \bigg]^{-2} V_{\rm surv}(z_{\rm c}),
\end{equation}
where $\overline{n}_{\rm g}(z)P_{\rm obs}(k,\mu,z)$ is the signal-to-noise, the comoving survey volume $V_{\rm surv}(z_{\rm c})$ with the redshift bin width $\Delta z$ is given by
\begin{equation}\label{eq:4.9}
V_{\rm surv}(z_{\rm c}) = \frac{4\pi}{3}\bigg[r\Big(z_{\rm c}+\frac{\Delta z}{2}\Big)^{3} - r\Big(z_{\rm c}-\frac{\Delta z}{2}\Big)^{3}  \bigg] \frac{\Omega_{\rm surv}}{\Omega_{\rm sky}},
\end{equation}
where $\Omega_{\rm surv}$ and $\Omega_{\rm sky}$ are respectively the survey area and the area of the full sky. Additionally, $\theta$ is the 8-dimensional parameter vector which consists of five cosmological parameters and three oscillation parameters,
\begin{equation}\label{eq:4.10}
\omega_{\rm c}=\Omega_{\rm c}h^2, \omega_{\rm b}=\Omega_{\rm b}h^2, h, n_{\rm s}, A_{\rm s}, A, \omega, \phi.
\end{equation}
The partial derivatives of $P_{\rm obs}(k, \mu, z_{\rm c})$ w.r.t. the cosmological parameters are calculated numerically using {the finite difference,}
\begin{equation}\label{eq:4.11}
\frac{\partial P_{\rm obs}(k, \mu, z_{\rm c})}{\partial \theta_{i}} = \frac{P_{\rm obs}({\theta^{\rm fid}_i} + \Delta \theta_i) - P_{\rm obs}({\theta^{\rm fid}_i} - \Delta \theta_i)}{2\Delta \theta_i},
\end{equation}
where $\Delta \theta_i$ is taken to be $10\%$ of the fiducial value of {$\theta^{\rm fid}_i$}, 
{though we have explicitly checked} that the partial derivative is insensitive to the size of $\Delta \theta_i$. By contrast, the partial derivatives w.r.t. the oscillation parameters can be calculated analytically due to the analytic form of the oscillations.

The Fisher matrices of \Baojiunew{the different} redshift bins are 
summed up to 
\Baojiu{get} a $8 \times 8$ 
matrix, 
\Yuhao{and then we can calculate the covariance matrix by taking the inverse of this Fisher matrix and the uncertainties of the parameters are given by the square roots of its diagonal elements.} Since we \Baojiunew{are} mainly 
\Baojiunew{interested in} the constraints on the oscillation parameters, we marginalise the cosmological parameters 
using the analytic marginalisation method 
{given by} \citet{taylor2010}, which marginalises the nuisance parameters and preserves the information about the target parameters. The marginalised Fisher matrix is 
\Baojiunew{given by}
\begin{equation}\label{eq:4.12}
F_{\alpha\beta}^{\rm M} = F_{\alpha\beta} - F_{\alpha m}F_{mn}^{-1}F_{n\beta},
\end{equation}
where the subscripts $\alpha$ and $\beta$ denote the target parameters, while the subscripts $m$ and $n$ denote the nuisance parameters. Finally, we get the uncertainties of the oscillation parameters from \Yuhao{the marginalised Fisher matrix}.

\subsection{Parameters used in the Fisher analysis}
\label{sec:3.3}

The parameters {used in the Fisher analysis, including those} associated with the specifications of the DESI-like survey \citep{aghamousa2016} are discussed here. 

We 
{start with} the most crucial parameter, 
the damping parameter $\zeta$ displayed in Table~\ref{tab2}, which depends not only on the redshifts but also on the halo number densities 
and {-- more importantly --} whether the reconstruction is applied. We only have values of $\zeta$ for four redshifts, i.e., $z=0$, $0.5$, $1$, $1.5$, and two different halo number densities, i.e., $n_{\rm halo}=1\times10^{-3}(h^{-1} \rm Mpc)^{-3}$ and $5\times10^{-4}(h^{-1} \rm Mpc)^{-3}$, but the forecasted number density achievable in the DESI-like survey varies over the redshift range, so the values of $\zeta$ may not apply to the entire redshift range. As a result, we cut off some high redshift bins which have the number density much smaller than $5\times10^{-4}(h^{-1} \rm Mpc)^{-3}$. We use a bilinear interpolation between the redshift and the number density to estimate an appropriate value of $\zeta$ for a given combination of the redshift and number density. For those the number density is larger than $1\times10^{-3}(h^{-1} \rm Mpc)^{-3}$ or smaller than $5\times10^{-4}(h^{-1} \rm Mpc)^{-3}$, we simply adopt the values of $\zeta$ for $n_{\rm halo}=1\times10^{-3}(h^{-1} \rm Mpc)^{-3}$ or $n_{\rm halo}=5\times10^{-4}(h^{-1} \rm Mpc)^{-3}$ instead. {In this work, we use different values of $\zeta$ for the different models as obtained using the fitting method described in Section \ref{sec:2.4}, and we will comment on this point again later.}

As we consider both emission line galaxies (ELGs) and luminous red galaxies (LRGs) in the DESI-like survey, which have different number densities and redshift distributions, different range of redshift bins is chosen for ELGs and LRGs in the Fisher analysis. After 
{throwing away} the redshift bins with very small number densities, we 
take the range of $z=(0.6-1.3)$ for ELGs and $z=(0.6-0.9)$ for LRGs, and the redshift bin width is by default $\Delta z=0.1$. In addition to the calculation of effective survey volume, by following the DESI-like survey, the fixed values of $\overline{n}_{\rm g}(z)P_{\rm obs}(0.14,0.6,z)$ are used for the signal-to-noise, two survey areas are considered including the expected survey area of 14,000 $\rm deg^2$ and 9,000 $\rm deg^2$ as the pessimistic case \citep{aghamousa2016}. As for the Finger-of-God factor, the linear halo bias for ELGs and LRGs is simply defined in terms of the growth factor via \citep{aghamousa2016}
\begin{equation}\label{eq:4.13}
b_{\rm ELG}(z)D(z) = 0.84
\quad\mathrm{and}\quad
b_{\rm LRG}(z)D(z) = 1.70.
\end{equation}
%
\begin{figure*}
	\includegraphics[width=2.0\columnwidth]{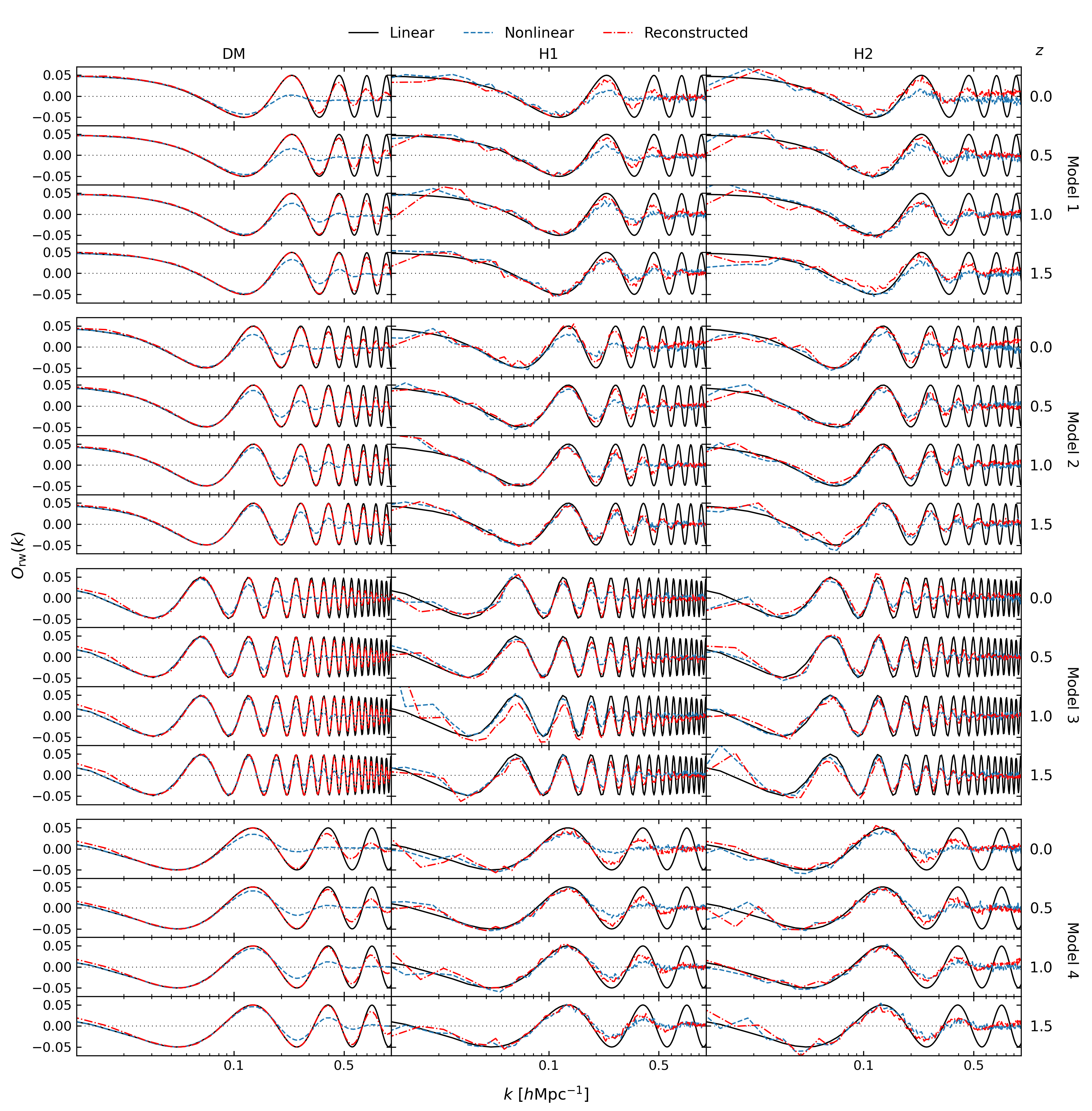}
    \caption{{[Colour Online]} 
    Comparisons among the linear (black solid line), nonlinear (blue dash-dotted line) and reconstructed 
    (red dashed line) {$\Yuhao{O}_{\rm rw}$}. The linear 
    {$\Yuhao{O}_{\rm rw}$} is measured from the initial conditions generated using 2LPTic, the nonlinear 
    {$\Yuhao{O}_{\rm rw}$} is measured from the output snapshots of the simulations, and the reconstructed 
    {$\Yuhao{O}_{\rm rw}$} is obtained from the 
    {reconstructed density field}. Each row represents one redshift $z$ which is shown on the right side. The three columns denote, respectively, {the results from} the dark matter particle catalogue DM and the halo catalogues H1 and H2. Every four 
    {rows} from the top down respectively belong to Model 1, Model 2, Model 3 \Yuhaonew{and Model 4}.}
    \label{fig:2}
\end{figure*}

\section{Results and discussion}
\label{sec:4}
In this section, we will first compare the linear, nonlinear and reconstructed 
$\Yuhao{O}_{\rm rw}$ measured 
{for all models and redshifts}. Then we 
{present} the results of the analytic fit to 
{more quantitatively demonstrate} the improvement 
by the reconstruction. Finally we show the results of the constraints on the oscillation parameters and give forecast for the DESI-like survey.

\subsection{Comparisons among wiggle spectra}
\label{sec:4.1}
In Fig.~\ref{fig:2}, we compare the results of the linear, nonlinear and reconstructed 
{$\Yuhao{O}_{\rm rw}(k)$} obtained from DM, H1 and H2 at the four redshifts for the \Yuhaonew{four} wiggled models. The black solid lines represent the linear 
{$\Yuhao{O}_{\rm rw}(k)$} obtained from the initial conditions of the simulations, which are equivalent to the primordial oscillatory features. The blue dashed lines represent the nonlinear 
{$\Yuhao{O}_{\rm rw}(k)$} obtained from the output snapshots of the simulations, which are also referred to as the unreconstructed 
{$\Yuhao{O}_{\rm rw}(k)$} for {convenience.}
It can be seen that the wiggles on small scales are gradually 
{damped} as the redshift decreases. The red dash-dotted lines represent the reconstructed 
{$\Yuhao{O}_{\rm rw}(k)$} obtained from the 
{reconstructed density field}, which 
{helps to partially retrieve} the damped wiggles.

The $\Yuhao{O}_{\rm rw}(k)$ results shown in the first column are obtained from DM, which exhibit some common characteristics for {all} three wiggled models. By comparing the unreconstructed
results with the linear-theory predictions, it can be seen that the scale at which the wiggles start to be weakened becomes larger as time progresses. 
Furthermore, the wiggles on scales $k \gtrsim 0.3 \ h \rm Mpc^{-1}$ are strongly damped at $z=0$, and so the recovery of the wiggles on scales $0.3 \lesssim k \lesssim 0.5 \ h \rm Mpc^{-1}$ would be an important objective of reconstruction. By comparing the reconstructed $\Yuhao{O}_{\rm rw}$ with the linear-theory prediction, we can see that, \Yuhao{while the reconstructed power spectrum is not exactly the same as the linear spectrum,} the reconstruction \Baojiunew{method to a certain extent helps retrieve} the initial oscillations on our interested scales, $0.05 \lesssim k \lesssim 0.5 \ h \rm Mpc^{-1}$. \Baojiunew{This agrees with the findings in \citet{shi2018}, which studied the performance of the same reconstruction method in dark matter reconstruction.}

The success of the reconstruction from the dark matter particles {is largely thanks to their high number density, which allows the late-time nonlinear density field to be accurately produced: in this sense, reconstruction from DM can be considered as an idealised case or an upper limit, which will be difficult to achieve in real observations. For a rough comparison, we have shown, in the middle and right columns of Fig.~\ref{fig:2},} the $\Yuhao{O}_{\rm rw}$ results obtained from {the two} halo catalogues, H1 and H2, {which have number densities similar to typical real galaxy catalogues}. {These results are less impressive} than those for the dark matter particles because of the much smaller halo number densities. Also due to the small halo number densities, {these results are noisier,} which in theory can be {made smoother} by having more realisations of simulations{, or equivalently a larger volume}.

By comparing the results of H1 and H2 for the same model, we find that there is no {significant difference} in the unreconstructed {$\Yuhao{O}_{\rm rw}(k)$} at the same redshift, {because the number densities of these two halo catalogues only differ by a factor of 2}. In most cases the reconstructed {$\Yuhao{O}_{\rm rw}$ results} of H1 seem {slightly} better compared to those of H2, as a result of the slightly larger \Baojiunew{halo} number density {in H1, though the difference is again insignificant visually.} We shall revisit this point when discussing the analytical fit in the next subsection. 
Comparing the results with and without reconstruction, it is clear that the former does lead to less damped and sharper oscillation features, confirming that reconstruction can indeed help to partially 
\Yuhaonew{retrieve} the damped wiggles. This recovery seems more substantial at lower redshifts than at higher redshifts, since at higher redshifts there is \Baojiunew{less} damping \Baojiunew{in the unreconstructed power spectra} to start with. At lower redshifts, on the other hand, reconstruction can even recover some of the wiggles at $k\sim0.5\ h \rm Mpc^{-1}$, where \Yuhao{the wiggles are strongly damped} in the unreconstructed case. We expect that this will help \Baojiunew{to improve the accuracy of the} measurements of wiggle parameters, especially in models with few wiggles at $k\lesssim 0.3\ h \rm Mpc^{-1}$ --- we will discuss this in the parameter fittings next\footnote{\Baojiunew{This is actually one of the motivations for our specific parameter choices in the feature models of Eq.~\eqref{eq:2.2}, because we are particularly interested in cases where there are not many wiggles at $k\lesssim0.3h\mathrm{Mpc}^{-1}$ to maximally show the power of reconstruction.}}.

\Yuhaonew{Finally, we notice that in rare cases, for example H1 at $z=1$ and H2 at $z=1.5$ for Model 3, the reconstructed $O_{\rm rw}$ seems to be poorer than the unreconstructed one. The exact cause of this is not clear, but we note that for these two cases the unreconstructed $O_{\rm rw}$ happens to be very noisy and deviate strongly from their theoretical values at large scales (a similar 'correlation' can be observed in certain other panels across Fig.~\ref{fig:2}, though to a lesser extent). It is possible that the halo power spectra in these cases have inaccurate amplitudes of the oscillations on large scales, which affect the reconstruction results. Given that in both H1 and H2 this only affects a particular snapshot and not all snapshots, we suspect that it is related to the 
{only one realisation per model} we have used. Further investigation of this issue will be left for future works with more simulation realisations.}

\begin{table*}
\centering
\caption{The best-fit parameters of $\omega$, $\phi$ and $\zeta$ and their {$95\%$} uncertainties for {the} four wiggled models {studied in this work}. \Yuhaonew{The values of $\omega$ and $\phi$ are respectively in the units of ${\rm Mpc}^{m}$ and $\pi$}, {and} their theoretical values are shown below the title of each model {on the top of the table}. DM denotes the dark matter particle catalogue, H1 
the halo catalogue with $n_{\rm halo}=1\times10^{-3}(h^{-1} \rm Mpc)^{-3}$ and H2 
the halo catalogue with $n_{\rm halo}=5\times10^{-4}(h^{-1} \rm Mpc)^{-3}$. Each group of six rows includes the unreconstructed and reconstructed cases for the same redshift.}
\label{tab2}
\begin{tabular}{c|c|c|c|c|c|c|c|c}
\hline
\multicolumn{3}{|c|}{} & \multicolumn{3}{|c|}{Model 1} & \multicolumn{3}{|c|}{\blue{Model 2}} \\ 
\multicolumn{3}{|c|}{} & \multicolumn{3}{|c|}{$\omega=40$, $\phi=0$} & \multicolumn{3}{|c|}{\blue{$\omega=70$, $\phi=0$}} \\ 
\hline
$z$ & & $\rm para$ & $\rm DM$ & $\rm H1$ & $\rm H2$ & \blue{$\rm DM$} & \blue{$\rm H1$} & \blue{$\rm H2$} \\ 
\hline
\multirow{6}{*}{$0.0$} & \multirow{3}{*}{unrec} &
$\omega$ & 
$33.0\pm3.4$ & $36.5\pm1.6$ & $36.2\pm1.9$ & 
\blue{$68.5\pm1.1$} & \blue{$69.0\pm1.7$} & \blue{$68.4\pm1.7$} \\
 & &
$\phi$ &
$0.09\pm0.08$ & $0.02\pm0.04$ & $0.04\pm0.05$ &
\blue{$0.00\pm0.03$} & \blue{$0.01\pm0.05$} & \blue{$0.00\pm0.04$} \\ 
 & & 
$\zeta$ &
$7.23\pm0.95$ & $6.21\pm0.44$ & $6.90\pm0.53$ &
\blue{$7.43\pm0.26$} & \blue{$6.95\pm0.40$} & \blue{$7.27\pm0.41$} \\ 
 & \multirow{3}{*}{rec} &
$\omega$ &
$40.0\pm0.1$ & $40.3\pm0.4$ & $40.8\pm1.1$ &
\blue{$70.1\pm0.1$} & \blue{$70.1\pm1.0$} & \blue{$69.9\pm0.9$} \\ 
 & & 
$\phi$ &
$0.00\pm0.01$ & $-0.02\pm0.02$ & $-0.04\pm0.04$ &
\blue{$-0.01\pm0.01$} & \blue{$0.00\pm0.05$} & \blue{$0.01\pm0.04$} \\ 
 & & 
$\zeta$ &
$2.05\pm0.05$ & $3.41\pm0.13$ & $4.07\pm0.32$ &
\blue{$2.06\pm0.03$} & \blue{$3.58\pm0.28$} & \blue{$4.00\pm0.25$} \\ 
\hline

\multirow{6}{*}{$0.5$} & \multirow{3}{*}{unrec} &
$\omega$ & 
$37.6\pm1.4$ & $37.3\pm1.1$ & $37.7\pm1.0$ &
\blue{$69.1\pm0.4$} & \blue{$68.0\pm0.9$} & \blue{$68.2\pm1.2$} \\
 & & 
$\phi$ &
$0.02\pm0.04$ & $0.03\pm0.03$ & $0.02\pm0.03$ &
\blue{$0.00\pm0.01$} & \blue{$0.05\pm0.03$} & \blue{$0.02\pm0.04$} \\ 
 & & 
$\zeta$ &
$5.53\pm0.40$ & $5.27\pm0.32$ & $5.71\pm0.29$ &
\blue{$5.88\pm0.10$} & \blue{$5.72\pm0.23$} & \blue{$6.18\pm0.30$} \\ 
 & \multirow{3}{*}{rec} &
$\omega$ &
$40.0\pm0.1$ & $40.1\pm0.3$ & $40.3\pm0.7$ &
\blue{$70.1\pm0.1$} & \blue{$70.1\pm0.4$} & \blue{$69.7\pm0.6$} \\ 
 & & 
$\phi$ &
$0.00\pm0.01$ & $-0.01\pm0.02$ & $-0.01\pm0.03$ &
\blue{$-0.01\pm0.01$} & \blue{$0.00\pm0.02$} & \blue{$0.01\pm0.03$} \\
 & & 
$\zeta$ &
$1.52\pm0.04$ & $3.17\pm0.10$ & $3.75\pm0.20$ &
\blue{$1.51\pm0.04$} & \blue{$3.12\pm0.10$} & \blue{$3.83\pm0.17$} \\
\hline

\multirow{6}{*}{$1.0$} & \multirow{3}{*}{unrec} &
$\omega$ &
$38.6\pm0.6$ & $38.0\pm0.7$ & $37.5\pm0.9$ &
\blue{$69.5\pm0.2$} & \blue{$68.7\pm0.8$} & \blue{$69.0\pm0.8$} \\
 & &
$\phi$ &
$0.01\pm0.02$ & $0.01\pm0.03$ & $0.01\pm0.03$ &
\blue{$0.00\pm0.01$} & \blue{$0.02\pm0.03$} & \blue{$0.00\pm0.03$} \\
 & & 
$\zeta$ &
$4.39\pm0.18$ & $4.38\pm0.22$ & $5.08\pm0.26$ &
\blue{$4.73\pm0.05$} & \blue{$4.90\pm0.21$} & \blue{$5.48\pm0.20$} \\
 & \multirow{3}{*}{rec} &
$\omega$ &
$40.0\pm0.1$ & $40.1\pm0.3$ & $39.7\pm0.7$ &
\blue{$70.0\pm0.1$} & \blue{$70.0\pm0.5$} & \blue{$70.0\pm0.7$} \\
 & & 
$\phi$ & 
$0.00\pm0.01$ & $-0.01\pm0.02$ & $0.01\pm0.03$ &
\blue{$0.00\pm0.01$} & \blue{$0.00\pm0.02$} & \blue{$-0.01\pm0.03$} \\
 & & 
$\zeta$ &
$1.09\pm0.04$ & $3.09\pm0.11$ & $3.56\pm0.20$ &
\blue{$1.11\pm0.03$} & \blue{$3.23\pm0.14$} & \blue{$3.81\pm0.20$} \\
\hline

\multirow{6}{*}{$1.5$} & \multirow{3}{*}{unrec} &
$\omega$ &
$39.3\pm0.3$ & $38.1\pm0.7$ & $36.5\pm0.8$ &
\blue{$69.6\pm0.1$} & \blue{$69.1\pm0.8$} & \blue{$68.2\pm1.0$} \\
 & & 
$\phi$ &
$0.00\pm0.01$ & $0.02\pm0.03$ & $0.07\pm0.03$ &
\blue{$0.00\pm0.01$} & \blue{$0.01\pm0.03$} & \blue{$0.03\pm0.03$} \\
 & & 
$\zeta$ &
$3.60\pm0.10$ & $4.07\pm0.20$ & $4.92\pm0.25$ &
\blue{$3.90\pm0.02$} & \blue{$4.43\pm0.21$} & \blue{$5.17\pm0.26$} \\ 
 & \multirow{3}{*}{rec} &
$\omega$ & 
$40.0\pm0.1$ & $39.6\pm0.5$ & $40.1\pm1.1$ &
\blue{$70.0\pm0.1$} & \blue{$70.2\pm0.4$} & \blue{$69.6\pm1.1$} \\
 & & 
$\phi$ & 
$0.00\pm0.01$ & $0.02\pm0.02$ & $-0.01\pm0.05$ &
\blue{$0.00\pm0.01$} & \blue{$-0.01\pm0.02$} & \blue{$0.01\pm0.05$} \\ 
 & & 
$\zeta$ &
$0.84\pm0.03$ & $3.08\pm0.15$ & $3.66\pm0.32$ &
\blue{$0.88\pm0.03$} & \blue{$3.24\pm0.12$} & \blue{$3.90\pm0.29$} \\
\hline

\multicolumn{3}{|c|}{} & \multicolumn{3}{|c|}{{Model 3}} & \multicolumn{3}{|c|}{\blue{Model 4}} \\ 
\multicolumn{3}{|c|}{} & \multicolumn{3}{|c|}{{$\omega=150$, $\phi=0$}} & \multicolumn{3}{|c|}{\blue{$\omega=28.9$, $\phi=0$}} \\ 
\hline
$z$ & & $\rm para$ & {$\rm DM$} & {$\rm H1$} & {$\rm H2$} & \blue{$\rm DM$} & \blue{$\rm H1$} & \blue{$\rm H2$} \\ 
\hline

\multirow{6}{*}{$0.0$} & \multirow{3}{*}{unrec} &
$\omega$ & 
{$149.3\pm0.4$} & {$151.2\pm2.0$} & {$151.6\pm2.1$} & 
\blue{$26.1\pm0.6$} & \blue{$27.2\pm1.0$} & \blue{$25.9\pm1.0$} \\
 & &
$\phi$ &
{$0.01\pm0.01$} & {$-0.04\pm0.05$} & {$-0.05\pm0.04$} &
\blue{$0.11\pm0.04$} & \blue{$0.06\pm0.06$} & \blue{$0.14\pm0.06$} \\ 
 & & 
$\zeta$ &
{$7.95\pm0.09$} & {$7.44\pm0.51$} & {$8.51\pm0.50$} &
\blue{$6.96\pm0.23$} & \blue{$6.62\pm0.35$} & \blue{$6.68\pm0.36$} \\ 
 & \multirow{3}{*}{rec} &
$\omega$ &
{$150.1\pm0.1$} & {$150.7\pm0.6$} & {$150.7\pm1.0$} &
\blue{$29.0\pm0.1$} & \blue{$29.5\pm0.4$} & \blue{$28.7\pm0.6$} \\ 
 & & 
$\phi$ &
{$-0.01\pm0.01$} & {$-0.04\pm0.03$} & {$-0.03\pm0.04$} &
\blue{$-0.01\pm0.01$} & \blue{$-0.05\pm0.03$} & \blue{$0.03\pm0.05$} \\ 
 & & 
$\zeta$ &
{$2.13\pm0.04$} & {$3.70\pm0.19$} & {$4.04\pm0.28$} &
\blue{$2.11\pm0.04$} & \blue{$3.41\pm0.14$} & \blue{$3.75\pm0.22$} \\ 
\hline

\multirow{6}{*}{$0.5$} & \multirow{3}{*}{unrec} &
$\omega$ & 
{$149.6\pm0.2$} & {$150.4\pm0.9$} & {$150.4\pm1.5$} & 
\blue{$27.2\pm0.2$} & \blue{$27.2\pm0.6$} & \blue{$26.2\pm0.7$} \\
 & &
$\phi$ &
{$0.01\pm0.01$} & {$-0.01\pm0.02$} & {$-0.02\pm0.03$} &
\blue{$0.07\pm0.02$} & \blue{$0.06\pm0.05$} & \blue{$0.12\pm0.05$} \\ 
 & & 
$\zeta$ &
{$6.30\pm0.05$} & {$6.02\pm0.23$} & {$6.85\pm0.39$} &
\blue{$5.33\pm0.08$} & \blue{$4.96\pm0.23$} & \blue{$5.58\pm0.26$} \\ 
 & \multirow{3}{*}{rec} &
$\omega$ &
{$150.1\pm0.1$} & {$150.0\pm0.5$} & {$150.1\pm0.8$} &
\blue{$29.0\pm0.1$} & \blue{$29.1\pm0.3$} & \blue{$29.3\pm0.4$} \\ 
 & & 
$\phi$ &
{$-0.01\pm0.01$} & {$-0.01\pm0.02$} & {$-0.02\pm0.03$} &
\blue{$-0.01\pm0.01$} & \blue{$-0.03\pm0.03$} & \blue{$-0.03\pm0.04$} \\ 
 & & 
$\zeta$ &
{$1.63\pm0.03$} & {$3.37\pm0.15$} & {$3.96\pm0.23$} &
\blue{$1.53\pm0.04$} & \blue{$3.04\pm0.11$} & \blue{$3.46\pm0.16$} \\ 
\hline

\multirow{6}{*}{$1.0$} & \multirow{3}{*}{unrec} &
$\omega$ & 
{$149.7\pm0.1$} & {$149.9\pm0.9$} & {$149.3\pm0.9$} & 
\blue{$27.7\pm0.1$} & \blue{$27.1\pm0.5$} & \blue{$26.7\pm0.5$} \\
 & &
$\phi$ &
{$0.01\pm0.01$} & {$0.00\pm0.03$} & {$0.02\pm0.03$} &
\blue{$0.05\pm0.01$} & \blue{$0.08\pm0.04$} & \blue{$0.10\pm0.03$} \\ 
 & & 
$\zeta$ &
{$5.08\pm0.03$} & {$5.29\pm0.25$} & {$5.95\pm0.24$} &
\blue{$4.19\pm0.03$} & \blue{$4.30\pm0.20$} & \blue{$4.99\pm0.17$} \\ 
 & \multirow{3}{*}{rec} &
$\omega$ &
{$150.1\pm0.1$} & {$150.6\pm1.7$} & {$149.9\pm0.6$} &
\blue{$29.0\pm0.1$} & \blue{$28.7\pm0.4$} & \blue{$28.3\pm0.4$} \\ 
 & & 
$\phi$ &
{$-0.01\pm0.01$} & {$-0.04\pm0.08$} & {$0.00\pm0.03$} &
\blue{$-0.01\pm0.01$} & \blue{$0.02\pm0.03$} & \blue{$0.04\pm0.04$} \\ 
 & & 
$\zeta$ &
{$1.27\pm0.03$} & {$3.31\pm0.52$} & {$3.85\pm0.17$} &
\blue{$1.12\pm0.03$} & \blue{$3.04\pm0.13$} & \blue{$3.55\pm0.15$} \\ 
\hline

\multirow{6}{*}{$1.5$} & \multirow{3}{*}{unrec} &
$\omega$ & 
{$149.7\pm0.1$} & {$149.6\pm0.8$} & {$149.6\pm0.9$} & 
\blue{$28.0\pm0.1$} & \blue{$27.4\pm0.6$} & \blue{$27.0\pm0.6$} \\
 & &
$\phi$ &
{$0.01\pm0.01$} & {$0.01\pm0.03$} & {$0.01\pm0.03$} &
\blue{$0.04\pm0.01$} & \blue{$0.05\pm0.05$} & \blue{$0.08\pm0.04$} \\ 
 & & 
$\zeta$ &
{$4.21\pm0.02$} & {$4.88\pm0.23$} & {$5.71\pm0.24$} &
\blue{$3.41\pm0.02$} & \blue{$3.92\pm0.22$} & \blue{$4.73\pm0.21$} \\ 
 & \multirow{3}{*}{rec} &
$\omega$ &
{$150.1\pm0.1$} & {$150.0\pm0.7$} & {$150.0\pm1.2$} &
\blue{$28.9\pm0.1$} & \blue{$29.0\pm0.4$} & \blue{$28.9\pm0.5$} \\ 
 & & 
$\phi$ &
{$-0.01\pm0.01$} & {$-0.01\pm0.03$} & {$0.00\pm0.05$} &
\blue{$0.00\pm0.01$} & \blue{$-0.02\pm0.04$} & \blue{$-0.01\pm0.04$} \\ 
 & & 
$\zeta$ &
{$1.07\pm0.03$} & {$3.38\pm0.20$} & {$4.07\pm0.34$} &
\blue{$0.83\pm0.02$} & \blue{$3.19\pm0.16$} & \blue{$3.65\pm0.17$} \\ 
\hline

\end{tabular}
\end{table*}

\begin{table*}
\centering
\caption{\Yuhao{The ratios of unreconstructed to reconstructed $\zeta$, \Yuhaonew{$\zeta_{\rm unrec}/\zeta_{\rm rec}$}, obtained from Table~\ref{tab2}, which can be used to describe the reconstruction efficiency, in all cases \Baojiunew{considered in Table \ref{tab2}.}}}
\label{tab3}
\begin{tabular}{c|c|c|c|c|c|c|c|c|c|c|c|c} 
\hline
& \multicolumn{3}{|c|}{Model 1} & \multicolumn{3}{|c|}{\blue{Model 2}} & \multicolumn{3}{|c|}{{Model 3}} & \multicolumn{3}{|c|}{\blue{Model 4}}\\
\hline
$z$ & $\rm DM$ & $\rm H1$ & $\rm H2$ & \blue{$\rm DM$} & \blue{$\rm H1$} & \blue{$\rm H2$} & {$\rm DM$} & {$\rm H1$} & {$\rm H2$} & \blue{$\rm DM$} & \blue{$\rm H1$} & \blue{$\rm H2$}\\
\hline
$0.0$ & $3.52$ & $1.82$ & $1.69$ & \blue{$3.61$} & \blue{$1.94$} & \blue{$1.82$} & {$3.73$} & {$2.01$} & {$2.11$} & \blue{$3.30$} & \blue{$1.94$} & \blue{$1.78$} \\
$0.5$ & $3.64$ & $1.66$ & $1.52$ & \blue{$3.89$} & \blue{$1.83$} & \blue{$1.61$} & {$3.87$} & {$1.79$} & {$1.73$} & \blue{$3.48$} & \blue{$1.63$} & \blue{$1.61$} \\
$1.0$ & $4.03$ & $1.42$ & $1.43$ & \blue{$4.26$} & \blue{$1.52$} & \blue{$1.44$} & {$4.00$} & {$1.60$} & {$1.54$} & \blue{$3.74$} & \blue{$1.41$} & \blue{$1.41$} \\
$1.5$ & $4.29$ & $1.32$ & $1.34$ & \blue{$4.43$} & \blue{$1.37$} & \blue{$1.33$} & {$3.93$} & {$1.44$} & {$1.40$} & \blue{$4.11$} & \blue{$1.23$} & \blue{$1.30$} \\
\hline
\end{tabular}
\end{table*}
\subsection{
{Wiggle parameter fitting}}
\label{sec:4.2}

The corresponding best-fit parameters of $\omega$, $\phi$ and $\zeta(z)$, 
{as well as} their uncertainties, are given in Table~\ref{tab2}, which assist the understanding from a quantitative perspective. \Yuhao{The relevant figures showing the analytic fit to the data can be found in the Appendix.} As mentioned before, we will mainly focus on the results of H1 and H2, and so the results of DM would be taken as a reference and not be discussed in detail. The three parameters are mainly determined by the remaining peaks in the wiggles. We shall first discuss the results of the damping parameter, followed by the oscillation parameters, and then combine them to clarify the improvement given by reconstruction. 

The damping parameter $\zeta$ effectively describes the extent of the \Baojiunew{damping} effects \Yuhao{caused by the gravitational nonlinearities\footnote{\Baojiunew{Redistribution of matter due to baryonic processes, such as stellar and black hole feedback, could also lead to damping effects to the power spectrum, but that is less relevant for the range of scales we are interested in \citep[some of the recent galaxy formation simulations, e.g.,][predict that this affects the matter power spectrum at $k\gtrsim1h\mathrm{Mpc}^{-1}$]{Schaye:2014tpa,Springel:2017tpz}.}}} and {characterises the suppression} of the primordial oscillations. It is {zero in the linear regime, such as} at \Baojiunew{the initial redshift} $z=49$, and gradually increases \Baojiu{at lower} redshifts as the structures become progressively more nonlinear \Baojiu{and consequently more information of the wiggles in the primordial power spectrum gets \Baojiunew{damped}}. Thus reconstruction has the aim to reduce $\zeta$ and retrieve the primordial oscillations. Table~\ref{tab2} shows that the reconstructed values of $\zeta$ are evidently smaller than the unreconstructed values in all cases. Apart from a few high-redshift ($z>1$) cases, the uncertainties of most cases are also reduced after reconstruction,
which {confirms} that the reconstruction \Baojiu{successfully retrieves} the damped wiggles to \Baojiu{an appreciable} extent. Specifically, by comparing the cases among different models but the same catalogues and redshifts, the {corresponding} values after reconstruction seem to be {nearly} independent of {the} model, which implies that the improvement on the {recovery} of the wiggles does not \Baojiunew{strongly} depend on the shape of the primordial oscillations\footnote{This makes sense given that the amplitude of the primordial oscillations is relatively small in this work, so that the \Baojiunew{effects of the} wiggles can be considered as small perturbations to the primordial \Baojiunew{and subsequently the evolved nonlinear} density field. Reconstruction, \Baojiunew{along with the reduction of $\zeta$ from the unreconstructed to the reconstructed cases that it leads to}, is sensitive to the overall distribution of matter}.

\Yuhao{\Baojiu{For a closer inspection, we show} the ratios of unreconstructed to reconstructed $\zeta$ in Table~\ref{tab3}, \Yuhaonew{$\zeta_{\rm unrec}/\zeta_{\rm rec}$}, which can be considered as an indicator of the reconstruction efficiency. We do this for all the cases \Baojiu{(models, tracer types and redshifts) listed in Table \ref{tab2}}. The \Baojiu{reconstruction} efficiency of halo catalogues H1 and H2 increases with decreasing redshift, which shows that reconstruction is more beneficial for lower redshifts ($z<1$). \Baojiu{This is to be expected, given that the halo density field is more nonlinear at low $z$ and so the unreconstructed $\zeta$ is significantly larger than at high $z$; on the other hand, the reconstructed $\zeta$ depends more mildly on $z$, so that the ratio \Yuhaonew{$\zeta_{\rm unrec}/\zeta_{\rm rec}$} increases with decreasing $z$.} Also, among the low-redshift ($z<1$) cases, the larger number density of H1 leads to higher efficiency when compared \Baojiu{with} H2 at the same redshift.} 
For the DM case, the trend is reversed, with the ratio between unreconstructed and reconstructed $\zeta$ values increasing with redshift. Here the behaviour is quite different from the halo cases, with the reconstructed $\zeta$ decreasing much faster with increasing redshift $z$. \Yuhaonew{We have checked (though not shown here) that the values of $\zeta_{\rm unrec}/\zeta_{\rm rec}$ for the primordial features studied here are broadly consistent with the reconstruction efficiency defined in the same way applied to the reconstruction of BAO wiggles in \citet[][which uses the same reconstruction method and similar tracer number density]{Birkin:2018nag}.}

Next, let us consider whether the \Baojiu{`sharpened'} wiggles after reconstruction can lead to more accurate measurements of the oscillation parameters $\omega$ and $\phi$. Regarding the oscillation frequency $\omega$, the reconstructed values of $\omega$ are much closer to the theoretical values than the unreconstructed values in all cases, which is especially evident at low redshifts. Except for a few high-redshift cases, the improvement on the uncertainties after reconstruction is evident in most cases as well. 
\Yuhaonew{The unreconstructed $\omega$ values of Model 2 and Model 3 appear to be closer to their theoretical values than in Model 1 and Model 4, which is probably because the former two models have more oscillation periods within the fitting range of scales than the latter two} (see the right panel of Fig.~\ref{fig:1}, or the blue lines in Fig.~\ref{fig:2}). After reconstruction, however, there is less clear difference among the four models, either in how close the reconstructed $\omega$ is to the theoretical value or in their uncertainties. Likewise, the difference between the best-fit reconstructed $\omega$ values in H1 and H2 is rather mild, although the uncertainties are generally smaller for the former catalogue. \Yuhaonew{Overall, the results indicate that reconstruction does indeed lead to a stronger improvement of the measurement of $\omega$ in Models 1 and 4, which have fewer visible peaks at $k\lesssim0.5h\mathrm{Mpc}^{-1}$.} 

The situation is quite different in {the case of} the oscillation phase $\phi$. \Yuhaonew{The unreconstructed values of $\phi$ in Model 1, Model 2 and Model 3 are determined very well in most cases, so the reconstructed values only show a little improvement on the unreconstructed $\phi$ even for low-redshift cases. However, for Model 4 the unreconstructed values largely deviate from the theoretical value in all cases, and the unreconstructed values of H2 deviate even further than those of H1 at the same redshift.} Although we can not exclude the possibility that this discrepancy is an effect caused by the particular simulation, since we have only one realisation for each model, we doubt this would be the cause, because the same random phases have been used to generate the ICs for all simulations. Instead, we suspect that this is more likely to be caused by the fact that $m\neq1$ in \Yuhaonew{Model 4}, which means that the oscillation pattern is more complicated and thus leads to a less accurate fitting of $\phi$. Regardless, based on the table, it seems that the reconstruction once again enables more accurate measurement of $\phi$, especially for H2 at low redshift. 

\Yuhaonew{When considering the results of all three parameters, it seems that the reconstruction is most useful at low redshifts, $z<1$, and Model 1 and Model 4 benefit more from it than Model 2 and Model 3 do. Although the peaks of Model 2 and Model 3 are better preserved after the cosmic evolution so that their reconstructed results are better than those of the other two models, the improvement is relatively limited{, suggesting that} the improvement depends not only on how clear-cut the reconstructed wiggles are, but also on how poorly the primordial wiggles are preserved before reconstruction. Overall, reconstruction seems more useful where the primordial wiggles are more damped\footnote{\Yuhaonew{This statement, of course, is based on the limited range of models we have studied here.}}. As we mentioned before, the wiggles on scales $k \gtrsim 0.3 \ h \rm Mpc^{-1}$ are strongly damped at $z=0$; Model 2 and Model 3 have exactly the first several original peaks outside this range of scales, so these peaks are effectively preserved at low redshift. By contrast, we designed Model 4 so that it has one original peak at the same position of the first peak of Model 2 which is effectively preserved, and its second peak is at the same position of the third peak of Model 2, which is strongly damped. Therefore, the primordial wiggles of Model 4 are preserved less well than those of Model 2, and this Model benefits more from the reconstruction. Similarly, Model 1 has two original peaks in the range $k\lesssim0.5\ h \rm Mpc^{-1}$: the first is at a smaller scale compared with the first peak of the other models and thus is not preserved as well as the first peak of the other models due to the stronger damping effect, while the second peak is completely damped. Therefore Model 1 and Model 4 both benefit from the reconstruction substantially more than Model 2 and Model 3.}

Additionally, the values of $\omega$ used in Model 1, Model 2 and Model 3 imply that the reconstruction method is not only effective at low frequency, such as $\sim40 \ \rm Mpc$, but also working well at {relatively higher frequency}, such as $\sim150 \ \rm Mpc$.

\begin{figure*}
	\includegraphics[width=2.0\columnwidth]{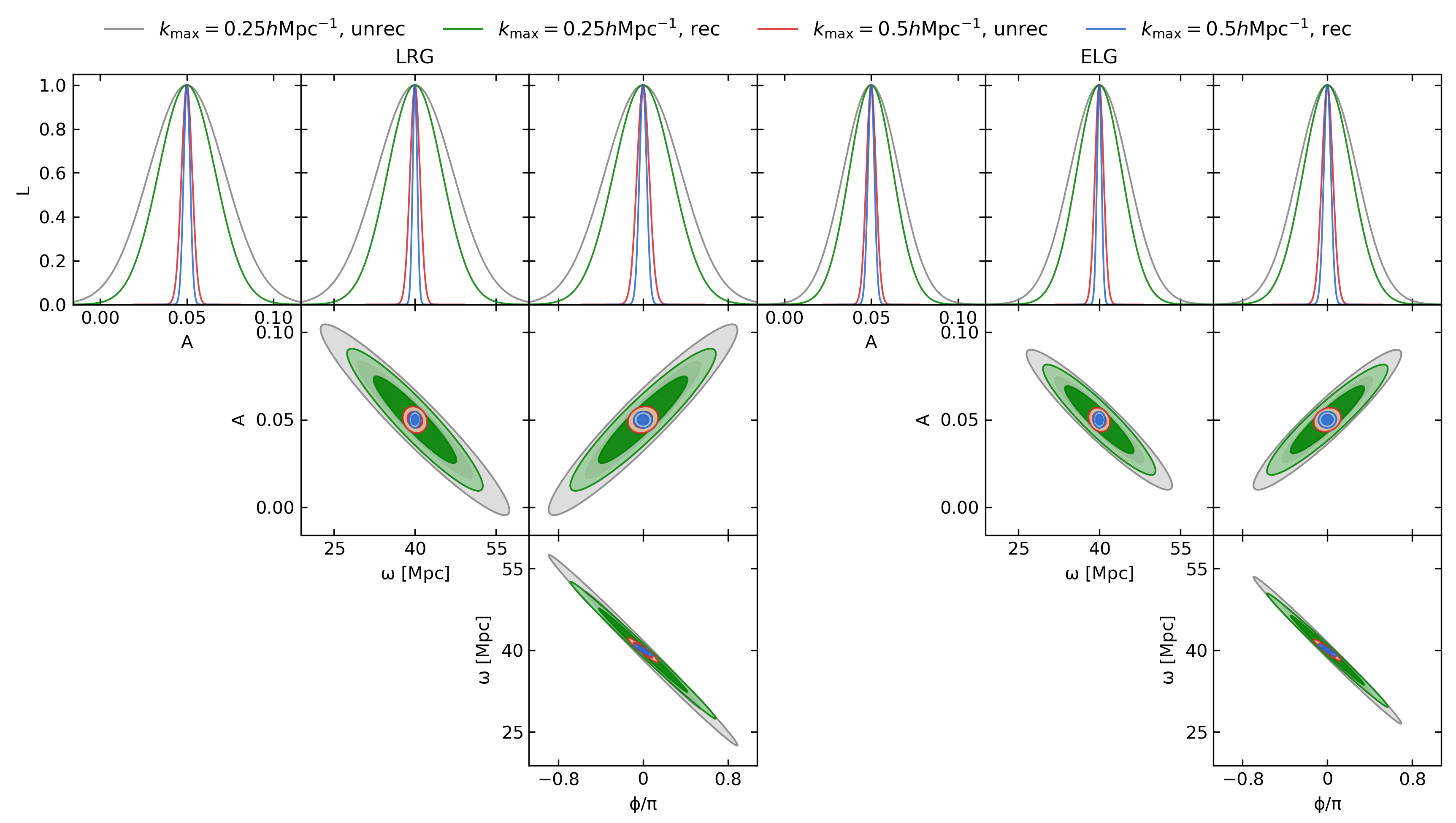}
    \caption{{[Colour Online]} {Forecasts of} constraints on the oscillatory feature parameters for 
    {a DESI-like} survey with 
    {a} survey area of $14,000$ $\rm deg^2$, 
    {for} the primordial oscillations of Model 1. The left side is for LRGs and the right side is for ELGs. The upper panels show the 1D marginalised 
    \Yuhao{posterior distributions}. The middle and lower panels show the marginalised $68\%$ and $95\%$ confidence contours for every two out of three feature parameters. The green and grey colours represent, respectively, the cases for $k_{\rm max}=0.25h \rm Mpc^{-1}$ with and without reconstruction, while {the} blue and red colours 
    represent the cases for $k_{\rm max}=0.5h \rm Mpc^{-1}$ with and without reconstruction. }
    \label{fig:6}
\end{figure*}
\begin{figure*}
	\includegraphics[width=2.0\columnwidth]{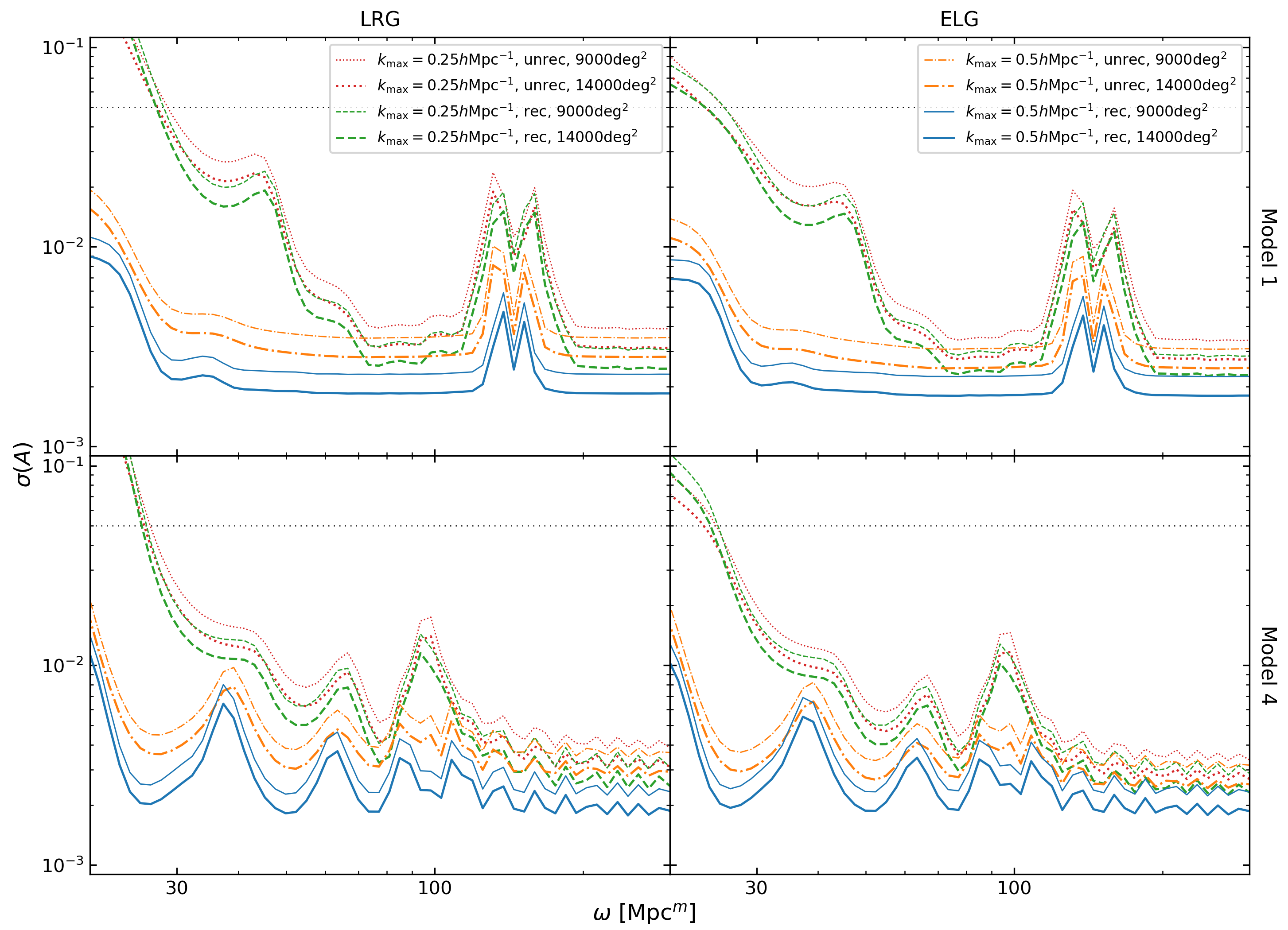}
    \caption{{[Colour Online]} Forecasts of the marginalised uncertainties of the oscillation amplitude $A$ as a function of the frequency $\omega$, 
    {for the} 
    \Baojiunew{two} models, \Baojiunew{Model 1 (top row) and \Yuhaonew{Model 4} (bottom row); the result for Model 2 and Model 3 are not shown here since the two models have the identical form of oscillations to that of Model 1}. The first column is for LRGs and the second column is for ELGs. 
    The dotted black lines mark the theoretical amplitudes {of the oscillations, $A=0.05$,} used in the forecasts. {The meanings of the different colours and line styles are indicated in the legends.} The same colours represent the cases with same $k_{\rm max}$ and same situation of reconstruction but different survey areas; the thick lines are for the survey area of $14,000$ $\rm deg^2$ and the thin lines are for $9,000$ $\rm deg^2$.}
    \label{fig:8}
\end{figure*}
\subsection{Constraints on oscillation parameters for DESI-like survey}
\label{sec:4.3}

\Yuhaonew{Since the four wiggled models have similar results of the constraints on the oscillation parameters, we shall take Model 1 as an example to illustrate and discuss how the reconstruction potentially improves the constraints in a real galaxy survey. 
Additionally, we also forecast how much the uncertainties of the feature amplitude can be reduced after reconstruction for the wiggled models.}

\Yuhaonew{Fig.~\ref{fig:6} shows the forecasted constraints on the oscillation parameters for a DESI-like survey with a survey area of 14,000 $\rm deg^2$, based on the primordial oscillations of Model 1. The marginalised posterior distribution of each parameter shown in the upper panels indicates that, without reconstruction, the case of $k_{\rm max} = 0.5 \ h \rm Mpc^{-1}$ (the red lines) give better constraints than the case with $k_{\rm max} = 0.25 \ h \rm Mpc^{-1}$ (grey), because in the former case more $k$ modes are included in the Fisher matrix and increase the accuracy of the constraints. Additionally, by comparing the cases with the same $k_{\rm max}$ (red versus blue, or grey versus green lines), we find that reconstruction leads to stronger constraints on the parameters, especially with $k_{\rm max} = 0.5 \ h \rm Mpc^{-1}$. This is because the oscillation wiggles on scales $k \gtrsim 0.25 \ h \rm Mpc^{-1}$ are heavily damped at low redshift without any reconstruction, while the reconstructed wiggles at $k = (0.25-0.5) \ h \rm Mpc^{-1}$ significantly contribute to the constraints. By contrast, since the peaks on scales $k \lesssim 0.25 \ h \rm Mpc^{-1}$ are preserved reasonably well, the reconstruction for $k_{\rm max} = 0.25 \ h \rm Mpc^{-1}$ does not lead to as much benefit as in the case of $k_{\rm max} = 0.5 \ h \rm Mpc^{-1}$. Furthermore, stronger constraints are shown for ELGs (right panels) compared with LRGs (left panels), because the former has more available redshift bins and larger number density for the same redshift bins.}

\Yuhaonew{In particular, every two out of three parameters show degeneracies in the confidence contours when $k_{\rm max} = 0.25 \ h \rm Mpc^{-1}$, though these degeneracies are broken and replaced with stronger constraints when $k_{\rm max} = 0.5 \ h \rm Mpc^{-1}$ in the $A$--$\omega$ and $A$--$\phi$ contours due to more $k$ modes included. By contrast, the $\omega$--$\phi$ contours keep the degeneracy which is a consequence caused by the oscillation model itself and by the fact that here we are trying to constrain both oscillatory frequency and phase over a limited range of $k$.} 

Lastly, {similar to previous works \citep{2019BAAS...51c..98S, Beutler:2019ojk, Ballardini:2019tuc},} we show the marginalised uncertainties of feature amplitude as a function of oscillatory frequency for 
our feature models in Fig.~\ref{fig:8} and discuss the implications of {the results}. Because Model 1, Model 2 and Model 3 have an identical form of oscillations and almost same damping parameters within the error bars, we only show the results of Model 1 and \Yuhaonew{Model 4} here.  

We consider Model 1 first. 
As expected, ELGs 
{place slightly} tighter constraints than LRGs due to their larger number densities {and redshift range}. The sharp peaks that appear at \Yuhaonew{$\omega \simeq 150 \ {\rm Mpc}$} are due to the degeneracy between the oscillatory features and the BAO {wiggles}. 
\Yuhaonew{We have tested that for $\omega \gtrsim 200 \ {\rm Mpc}$ the uncertainties almost stay as a constant, and so we have cut off the figure at $\omega^m=300\ {\rm Mpc}^m$}. For smaller $\omega$, things are complicated {and behave differently} for different $k_{\rm max}$. For $k_{\rm max} = 0.25 \ h \rm Mpc^{-1}$ we can see {an increase} in the uncertainties at \Yuhaonew{$\omega \lesssim 70 \ {\rm Mpc}$}, while a similar {increase} starts to appear at even smaller \Yuhaonew{$\omega$ -- $30 \ {\rm Mpc}$} -- for $k_{\rm max} = 0.5 \ h \rm Mpc^{-1}$. Thus larger $k_{\rm max}$ has an {extra} advantage of {significantly} reducing the uncertainties for small $\omega${, in addition to giving more stringent constraints (everything else the same) for all $\omega$ overall}. {By comparing} the pairs of {curves with} the same colours, i.e., {the} same cases {($k_{\rm max}$ and reconstructed vs.~unreconstructed)} but different survey areas, {we find that, as expected, a} larger survey area always {gives} better constraints.

Most interestingly, everything else equal, performing \Yuhao{the nonlinear} reconstruction can significantly reduce the uncertainties of $A$. As an example, for large values of $\omega$, in the case of $k_{\rm max}=0.5\ h \rm Mpc^{-1}$ and a survey area equal to $14,000~{\rm deg}^2$, reconstruction reduces $\sigma(A)$ from $\sim0.003$ to $\sim0.002$, and this improvement is stronger than not performing reconstruction, but instead going from $9,000$ to $14,000$ ${\rm deg}^2$ with $k_{\rm max}$ fixed to $0.25$ or $0.5\ h \rm Mpc^{-1}$, or increasing $k_{\rm max}$ from $0.25$ to $0.5\ h \rm Mpc^{-1}$ keeping the survey area fixed to either $9,000$ or $14,000$ ${\rm deg}^2$. A similarly good improvement can be seen with $k_{\rm max}=0.25\ h \rm Mpc^{-1}$ or survey area equal to $9,000$ ${\rm deg}^2$, when doing reconstruction. In certain cases, e.g., the large-$\omega$ regime of the lower panels of Fig.~\ref{fig:8}, reconstruction with $k_{\rm max}=0.25\ h \rm Mpc^{-1}$ and a survey area equal to $9,000$ ${\rm deg}^2$ (the thin green dashed line) can lead to comparable constraints to not doing reconstruction but with $k_{\rm max}=0.5\ h \rm Mpc^{-1}$ and a survey area equal to $14,000$ ${\rm deg}^2$ (the thick orange dot-dashed line). 
\Yuhao{Given that increasing survey area is not always possible due to the finite sky area, but increasing $k_{\rm max}$ in analyses for these primordial feature models is comparably \Baojiu{more straightforward} \citep{Beutler:2019ojk}, combining an increase in $k_{\rm max}$ with \Yuhao{nonlinear} reconstruction can \Baojiunew{be a potentially promising way to} obtain even stronger constraints on the feature parameters, and help to maximise the scientific return of future survey data.}


The behaviour of \Yuhaonew{Model 4} is similar to that of Model 1, e.g., both the absolute and the relative heights of the different curves, as well as their shapes are the same as before. There are, however, some notable differences, e.g., the main peaks in $\sigma(A)$ in \Yuhaonew{Model 4} are at slightly different values of $\omega$ from the other models, and the curves are also less smooth. As mentioned above, the bump (which has the structure of a double peak) of $\sigma(A)$ for Model 1 is related to the BAO peak in the matter/galaxy correlation function, which is at \Yuhaonew{$\simeq150 \ {\rm Mpc}$}. The primordial wiggles of Model 1, in configuration space, correspond to a spike at matter or halo separation $r=\omega$. When \Yuhaonew{$\omega\gg150 \ {\rm Mpc}$}, the BAO and primordial peaks are separated afar and thus the former does not affect the accuracy of the measurement for the latter. As $\omega$ approaches \Yuhaonew{$150 \ {\rm Mpc}$} from above, the BAO and primordial peaks start to `interfere', leading to changes of both the amplitude and shape of the latter, making it harder to measure its parameters accurately. We speculate that the dip --- which causes the double-peak structure in $\sigma(A)$ for Model 1 --- is due to the fact that, when the primordial peak does not coincide well with the centre of the (rather wide) BAO peak, its shape can be affected in an asymmetric manner, making the measurement of its parameters even more inaccurate. In contrast, the structure of the primordial wiggles in \Yuhaonew{Model 4} is more complicated in configuration space, because $m\ne1$ in Eq.~\eqref{eq:2.2}, which can cause the differences in the \Yuhaonew{units of $\omega$ and other} fine details of $\sigma(A)$ between this and the other models. 

\section{Conclusions}
\label{sec:5}

In this paper, we have investigated the effect of \Baojiunew{a nonlinear density} reconstruction method on retrieving {hypothetical} oscillatory features in the primordial power spectrum which are \Yuhao{significantly damped} on small scales in the late-time Universe due to 
cosmological \Baojiunew{structure formation.} 

{We considered} \Yuhaonew{four} different oscillatory features {which are added} to a simple power-law type primordial power spectrum, {for which} we ran N-body simulations and identified dark matter halo catalogues at a number of redshifts. We reconstructed the initial density fields from the particle data and halo catalogues with two different number densities. Finally, we compared the fitted feature parameters from \Baojiunew{the power spectra of} the unreconstructed and reconstructed {density fields,} to identify the improvement by reconstruction. We showed that \Yuhao{nonlinear} reconstruction \Baojiunew{can effectively help to retrieve} the damped wiggles \Yuhaonew{with a range of frequencies between $40 \ \rm Mpc$ and $150 \ \rm Mpc$} --- not only does it lead to less biased best-fit values of the feature parameters, but it also substantially shrinks the measurement uncertainty. The improvement was especially {strong where the primordial features have been 
\Baojiunew{less well preserved pre-reconstruction} to start with, such as at $z<1$.}

In order to forecast the constraints on the feature parameters from a DESI-like galaxy survey, we modelled the observed broadband galaxy power spectrum based on the {\sc halofit} \Baojiunew{prediction of the nonlinear matter power spectrum} with the addition of oscillatory features \Yuhao{studied in this work}, and then used the analytic marginalised Fisher matrix to calculate the \Baojiunew{expected} constraints on the oscillation parameters \Baojiunew{using} the specifications of DESI LRGs and ELGs. We found that \Yuhao{nonlinear} reconstruction led to more robust constraints on the oscillation parameters, with the equivalent effects of enlarging the survey area (but at a much smaller cost) and/or increasing the $k$ range.

While \Yuhao{nonlinear} reconstruction 
\Baojiu{has been proposed to be} used in improving the measurement of the BAO scale \citep[e.g.,][]{Wang:2017jeq}, and hence the determination of the expansion rate \Baojiunew{of the Universe} and \Baojiu{hence the} properties of dark energy, this work has demonstrated that similar applications are possible in other cases where certain features in matter clustering are present\Baojiunew{, following the spirit of earlier works such as \citet{Beutler:2019ojk}}. This is particularly true if these features are in the mildly nonlinear regime, $0.1\lesssim{k/(h{\rm Mpc}^{-1})}\lesssim0.5$, since this range of scales is what the nonlinear reconstruction method used here helps most: on even larger scales the benefit of reconstruction is insignificant, while on further smaller scales reconstruction will not help much. 

{The methodology exemplified in this paper assumes that we know the functional form of the primordial features {\it a priori} --- this is how we forecasted constraints on the oscillation amplitude $A$. However, the reconstruction step is completely independent of any assumption of a particular primordial feature, and hence any method developed for detecting general features from the matter clustering should apply to and benefit from the reconstructed density field.}

As a first step, the present study is based on various simplifications, and we discuss a couple here which can be improved in the future. The first is related to the post-reconstruction damping parameter $\zeta$. As we have discussed, $\zeta$ characterises the damping of the primordial features, and a smaller $\zeta$ means that the reconstruction has done a better job. \Yuhaonew{Due to the limited number of simulations carried out in this work (one realisation per model), shot noise will impact the estimated reconstruction efficiency. This could be improved by increasing the number of simulations and more studies are needed in the future.}

{The second is related to the modelling of redshift-space distortions (RSD), for which we have adopted a simplistic prescription and well pushed beyond the limit (e.g., $k\simeq0.5~h{\rm Mpc}^{-1}$) where it is expected to work. This is not an issue for a forecast work, but for constraints using real data it should be treated more carefully. The reconstruction method here has been extended to remove RSD from observed galaxy catalogues \citep{Wang:2019zuq}, though that is unlikely to work reliably at $k$ as large as $\simeq0.5~h{\rm Mpc}^{-1}$. Of course, we can always cut $k_{\rm max}$ to something that we are comfortable with. However, as mentioned above, if we would like to take maximum benefit from reconstruction, it is likely that we need to go substantially beyond $k\simeq0.1~h{\rm Mpc}^{-1}$. This can be achieved, for example, by using emulators of redshift-space galaxy or halo clustering \citep[see, e.g.,][]{Zhai:2018plk,Kobayashi:2020zsw}; actually, as long as the primordial oscillations are weak (as implied by current null detections), one might assume that their presence has little or negligible impact on RSD.}

{The ultimate objective, of course, is to apply this method to real observation data from future galaxy surveys such as Euclid and DESI. For this, the above-mentioned improvements, amongst many others, would need to be done properly. These will be left for future works, in which we plan to carry out updated forecasts for these surveys and eventually real constraints.}

\section*{Acknowledgements}

We thank collaborators within Euclid and DESI for various discussions while this project was going on. YL thanks Robert Smith for his support during this project. BL is supported by the European Research Council through ERC Starting Grant ERC-StG-716532-PUNCA, and the Science Technology Facilities Council (STFC) through ST/T000244/1 and ST/P000541/1. 
HMZ is supported by the Natural Sciences and Engineering Research Council of Canada (NSERC) [funding reference number CITA 490888-16]. This work used the DiRAC@Durham facility managed by the Institute for Computational Cosmology on behalf of the STFC DiRAC HPC Facility (www.dirac.ac.uk). The equipment was funded by BEIS capital funding via STFC capital grants ST/K00042X/1, ST/P002293/1, ST/R002371/1 and ST/S002502/1, Durham University and STFC operations grant ST/R000832/1. DiRAC is part of the National e-Infrastructure.


\section*{Data Availability}
 
Simulation data used in this work can be made available upon request to the authors. 
 


\bibliographystyle{mnras}
\bibliography{reference} 



\appendix

\section{Results of wiggle fitting}
\label{appendixA}
\begin{figure*}
	\includegraphics[width=2.0\columnwidth]{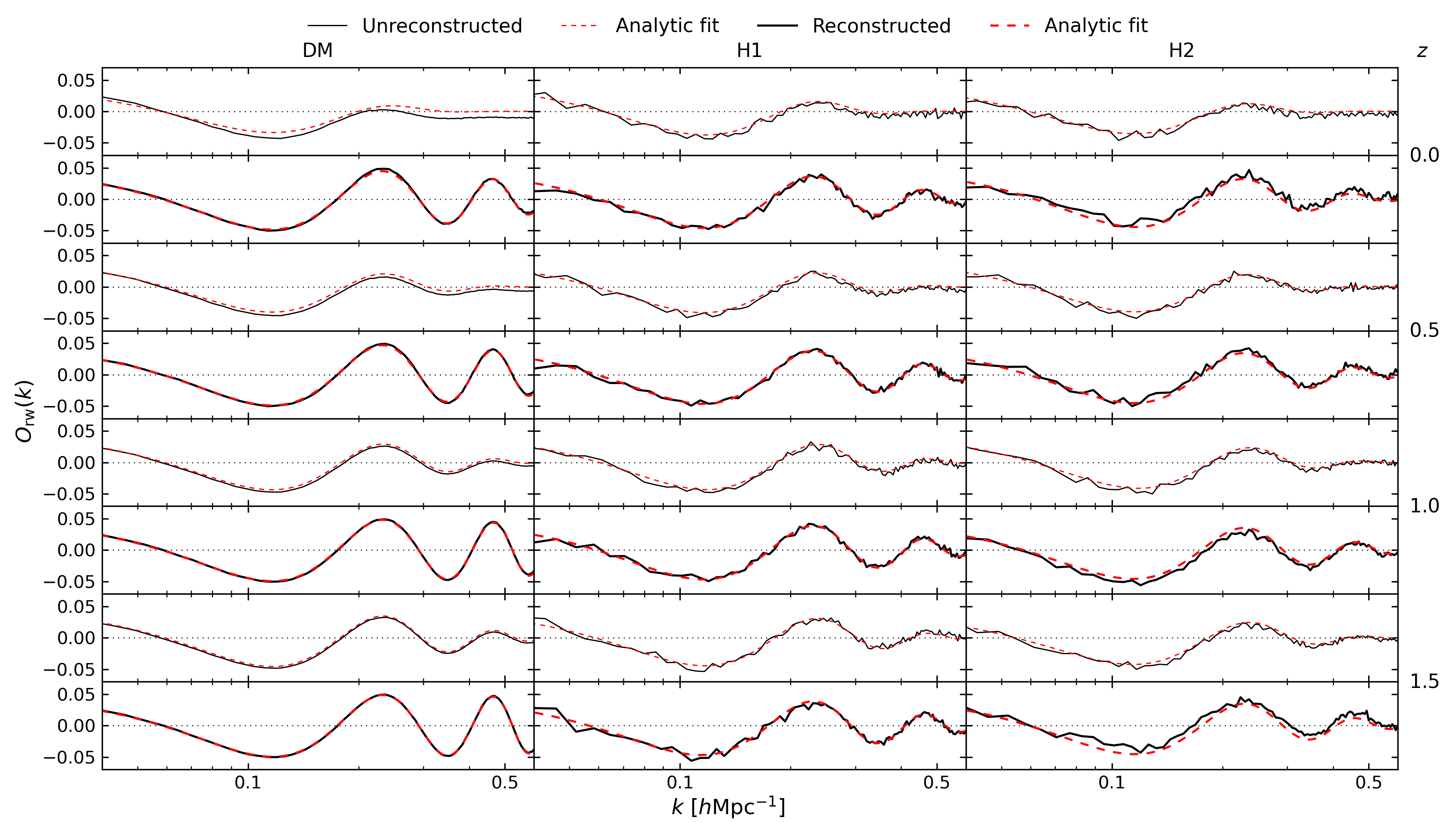}
    \caption{{[Colour Online]} The analytic fit to the unreconstructed and reconstructed 
    {$\Yuhao{O}_{\rm rw}$} for Model 1. The black solid lines represent the measured 
    {$\Yuhao{O}_{\rm rw}$} and the red dashed lines represent the fitting curves given by the analytic model{, Eq.~\eqref{eq:2.4}}. The thin lines are for the unreconstructed cases and the thick lines are for the reconstructed cases. The three columns {from left to right} respectively denote the dark matter particle catalogue DM, and the halo catalogues H1 and H2. Every two rows from the top down represent the same redshift shown on the right side. {In each group of two rows, the upper one is for the unreconstructed, and the lower one for the reconstructed, case.}}
    \label{fig:3}
\end{figure*}

\begin{figure*}
	\includegraphics[width=2.0\columnwidth]{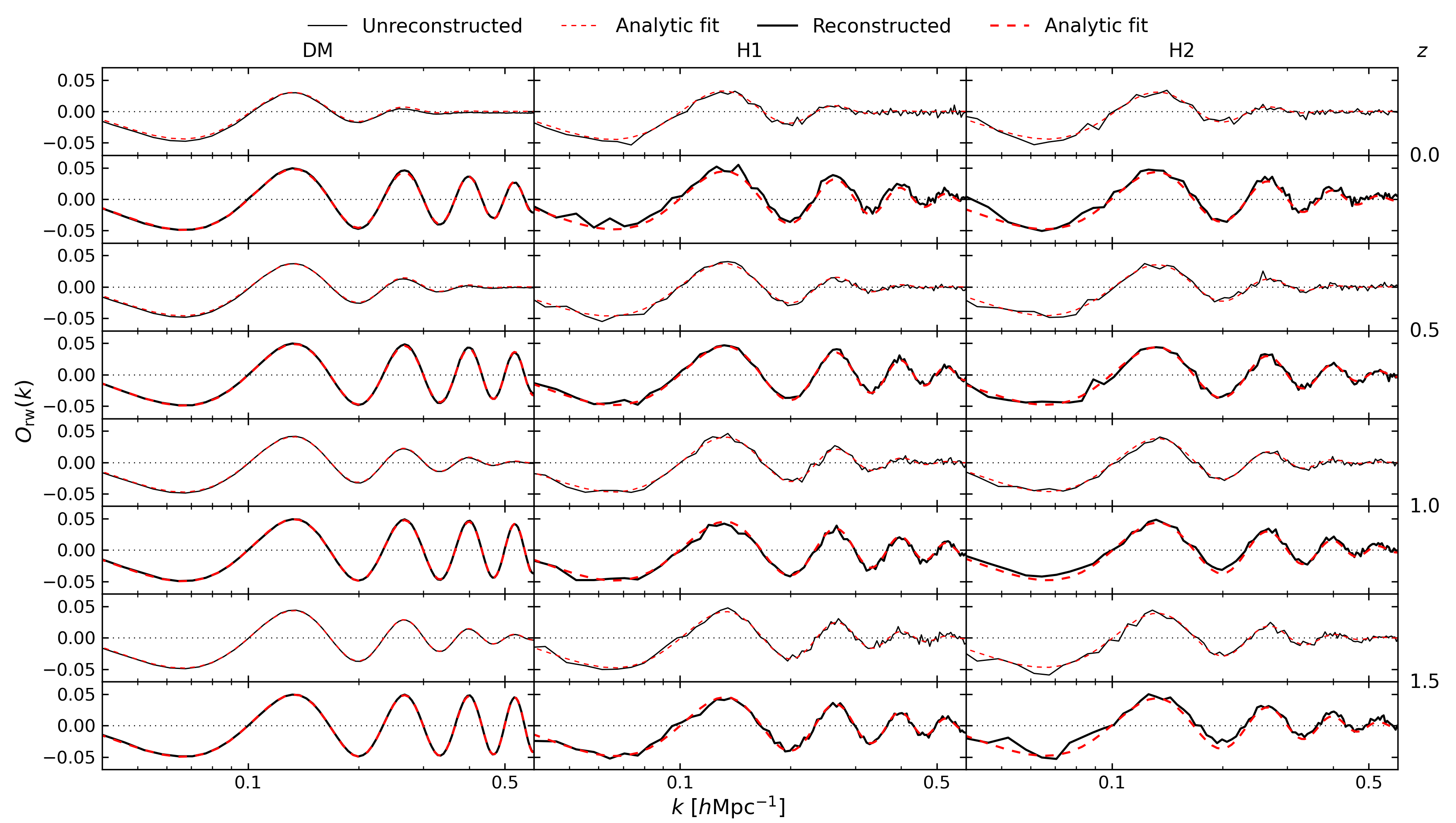}
    \caption{{[Colour Online]} The same as Fig.~\ref{fig:3} but for Model 2.}
    \label{fig:4}
\end{figure*}

\begin{figure*}
	\includegraphics[width=2.0\columnwidth]{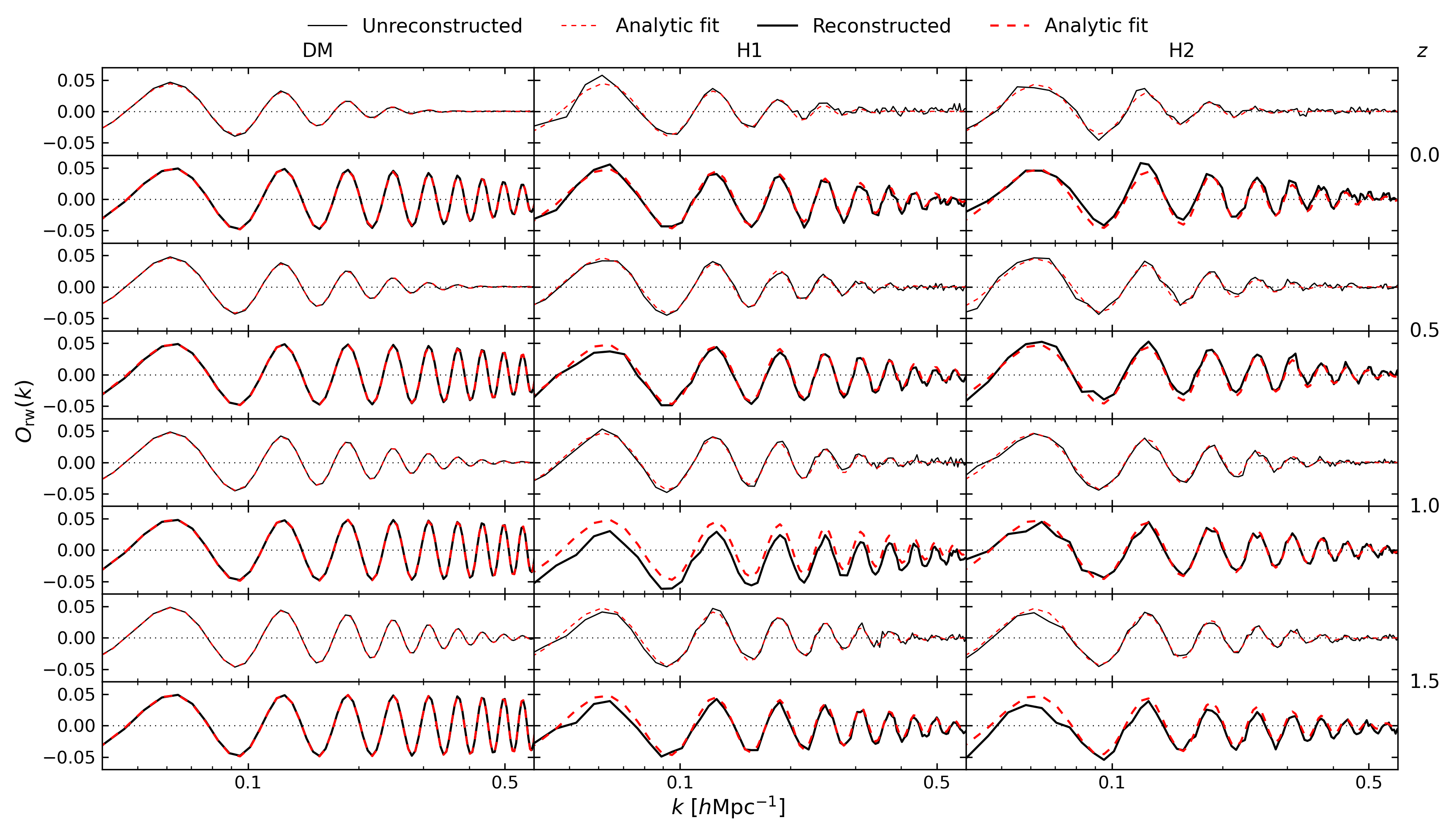}
    \caption{{[Colour Online]} The same as Fig.~\ref{fig:3} but for Model 3.}
    \label{fig:5}
\end{figure*}

\begin{figure*}
	\includegraphics[width=2.0\columnwidth]{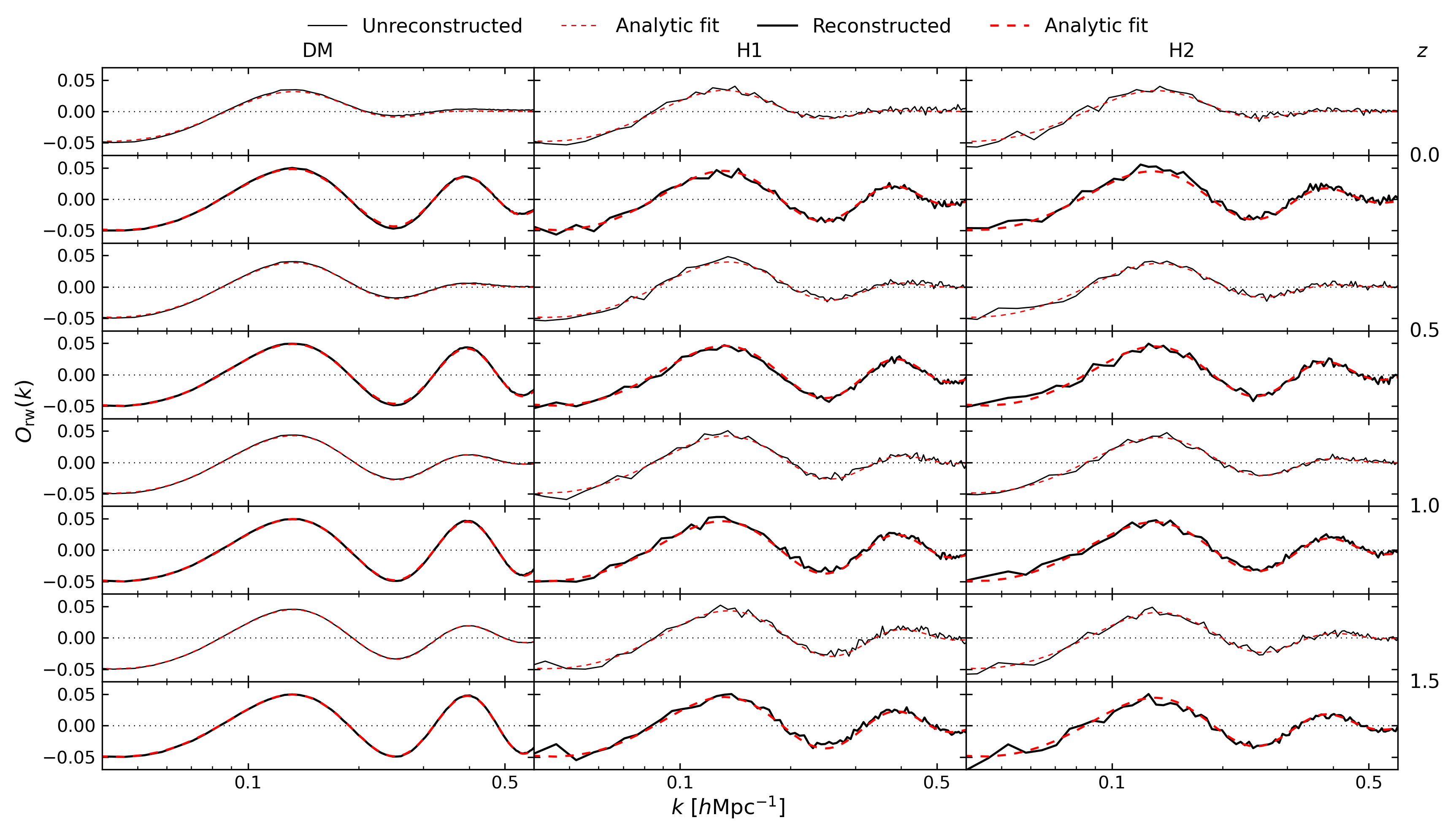}
    \caption{{[Colour Online]} The same as Fig.~\ref{fig:3} but for Model 4.}
    \label{fig:9}
\end{figure*}

{Figs.~\ref{fig:3}, \ref{fig:4}, \ref{fig:5} and \ref{fig:9} show,} respectively, 
the results of the analytic fit to {the} unreconstructed and reconstructed 
{$\Yuhao{O}_{\rm rw}$ results} for {the} \Yuhaonew{four} models {studied in this work}. It can be seen that, in most cases, 
{the} analytic model {Eq.~\eqref{eq:2.4}, with a Gaussian damping function characterised by the parameter $\zeta(z)$,} fits the 
pre- and post-reconstruction {data} very well. 


\bsp	
\label{lastpage}
\end{document}